\documentclass[aps,pra,twocolumn,superscriptaddress,raggedbottom,showpacs,floatfix,nofootinbib,preprintnumbers]{revtex4-2}
\usepackage{amsmath,amsfonts,amssymb,amsthm,graphics}
\usepackage[next]{inputenc}
\usepackage[colorlinks=true,citecolor=blue,linkcolor=blue]{hyperref}
\usepackage{bbm}
\usepackage{booktabs}
\usepackage{multirow}
\usepackage{hhline}
\usepackage{comment}
\usepackage{float}
\usepackage{latexsym}
\usepackage[dvipsnames]{xcolor}
\usepackage{graphicx}
\usepackage{epstopdf}
\usepackage{bm}% bold math
\usepackage{verbatim}
\usepackage{physics}
\usepackage{mathrsfs}
\usepackage{tikz}
\usepackage{chemfig}
\usepackage{pifont}% http://ctan.org/pkg/pifont
\usepackage{manfnt}
\usepackage{mathtools}
\usepackage{soul}

\usepgflibrary{patterns} 
\usepgflibrary[patterns] 
\usetikzlibrary{patterns} 
\usetikzlibrary[patterns]
\usetikzlibrary{shapes,shadows,arrows,positioning,graphs}

\newcommand{\mch}{\mathcal{H}}

\newcommand{\updiag}{
 \raisebox{-2pt}{\includegraphics[height=2.1ex]{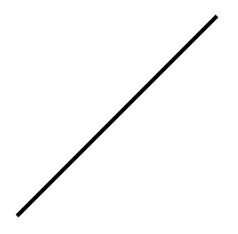}}}

\newcommand{\upbreak}{
 \raisebox{-2pt}{\includegraphics[height=2.5ex]{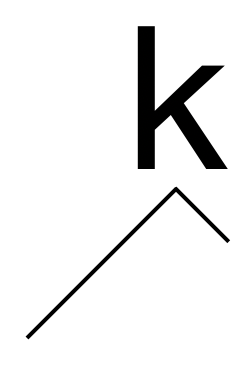}}}

\newcommand{\downdiag}{
 \raisebox{-2pt}{\includegraphics[height=2.1ex]{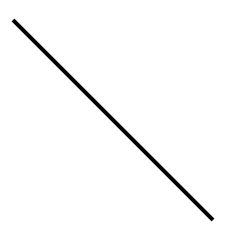}}}

\newcommand{\downbreak}{
 \raisebox{-2pt}{\includegraphics[height=2.5ex]{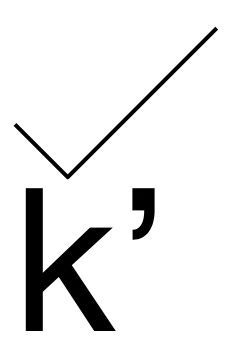}}}

\newcommand{\leftframe}{\mathord{\raisebox{-3pt}{\begin{tikzpicture}
\draw (0,-0.2) -- (0,0.2);
\draw[->] (0,0) -- (0.2,0);
\end{tikzpicture}}}}

\newcommand{\rightframe}{\mathord{\raisebox{-3pt}{\begin{tikzpicture}
\draw (0.2,-0.2) -- (0.2,0.2);
\draw[->] (0.2,0) -- (0,0);
\end{tikzpicture}}}}

\newcommand{\upframe}{\mathord{\raisebox{0pt}{\begin{tikzpicture}
\draw (-0.2,0) -- (0.2,0);
\draw[->] (0,0) -- (0,0.2);
\end{tikzpicture}}}}

\newcommand{\downframe}{\mathord{\raisebox{-2pt}{\begin{tikzpicture}
\draw (-0.2,0) -- (0.2,0);
\draw[->] (0,0) -- (0,-0.2);
\end{tikzpicture}}}}
   
\newtheorem{theorem}{Theorem}%[section]

\begin{document}

\title{2D Hamiltonians with exotic bipartite and topological entanglement}
\author{Shankar Balasubramanian}
\affiliation{Center for Theoretical Physics, Massachusetts Institute of Technology, Cambridge, MA 02139, USA}
\affiliation{Department of Physics, Massachusetts Institute of Technology, Cambridge, MA 02139, USA}
\author{Ethan Lake}
\affiliation{Department of Physics, Massachusetts Institute of Technology, Cambridge, MA 02139, USA}
\author{Soonwon Choi}
\affiliation{Center for Theoretical Physics, Massachusetts Institute of Technology, Cambridge, MA 02139, USA}
\affiliation{Department of Physics, Massachusetts Institute of Technology, Cambridge, MA 02139, USA}

\preprint{MIT-CTP/5546}

\begin{abstract}
We present a class of exactly solvable 2D models whose ground states violate conventional beliefs about entanglement scaling in quantum matter. 
These beliefs are (i) that area law entanglement scaling originates from local correlations  proximate to the boundary of the entanglement cut, and (ii) that ground state entanglement in 2D Hamiltonians cannot violate area law scaling by more than a multiplicative logarithmic factor.
We explicitly present two classes of models defined by local, translation-invariant Hamiltonians, whose ground states can be exactly written as weighted superpositions of framed loop configurations.  
The first class of models exhibits area-law scaling, but of an intrinsically nonlocal origin so that the topological entanglement entropy scales with subsystem sizes.
The second class of models has a rich ground state phase diagram that includes a phase exhibiting volume law entanglement. 
\end{abstract}

\maketitle

\emph{Introduction} --- 
Entanglement is one of the unique features that distinguish quantum systems from classical ones. 
Thus, it is natural that understanding entanglement structure has become a central tool in studying strongly correlated quantum matter. 
For ground states of gapped systems, it is widely believed that the entanglement entropy of a subregion $R$ scales linearly in the size of its boundary area $\partial R$ \cite{RevModPhys.82.277}. This phenomenon is dubbed the \emph{area law}, and has been proven in 1D \cite{hastings} and for frustration-free models in 2D \cite{anshu2022area}.  Sufficient conditions for area law scaling include ground state correlation decay and a sub-exponential number of states with vanishing energy density \cite{Masanes}.  Even for gapless systems, the entanglement entropy is expected to follow area-law scaling up to a universal logarithmic correction factor \cite{calabrese2004entanglement, calabrese2009entanglement}, which in 2D can be additive \cite{fradkinmoore} or multiplicative \cite{gioev2006entanglement}.

Entanglement gives fundamental insights into the classification of phases of matter. Within the purview of gapped phases, entanglement-based metrics can be used to study topological states, which are said to possess long-range entanglement.  In particular, the topological entanglement entropy (TEE)~\cite{KitaevPreskill, LevinWen} computes the subleading term $\gamma$ of the entanglement scaling $S(R) = \alpha |\partial R| - \gamma$. The TEE captures intrinsically long-ranged entanglement, since nonvanishing $\gamma$ for a gapped system implies the ground state is topologically ordered.  Recently, entanglement-based metrics were proposed to extract the chiral central charge of a ground state wavefunction~\cite{PhysRevLett.128.176402, kim}, which has led to an ongoing effort seeking to use entanglement-based metrics to classify topological phases of matter assuming a set of entropic axioms \cite{shi1, shi2, shi3, shi4, fan1, fan2}.  It is therefore important to develop a full understanding of the different kinds of entanglement that can arise in ground states of quantum many-body systems. 

In this work, we present exactly solvable 2D models whose entanglement properties defy conventional expectations.  These models are geometrically local, translationally invariant, and possess unique ground states.  
They can be viewed as 2D generalizations of the colored Motzkin spin model in 1D ~\cite{movassagh} (the uncolored version was introduced in~\cite{bravyi}), whose ground state violates the area law by a power law.  The first model we construct is a certain kind of loop model with area law entanglement scaling. The unusual aspect of this model is that the area law is due to \emph{intrinsically nonlocal} correlations.  This nonlocality reflects itself in the TEE,  which exhibits an area law term (as opposed to being an $O(1)$ constant).
We call this phenomenon \emph{anomalous} TEE. An anomalous TEE occurs in both the Kitaev-Preskill and Levin-Wen prescriptions.

We then modify our loop model by introducing additional degrees of freedom which decorate the loops with 1D Motzkin chains, resulting in volume-law entanglement scaling in the ground state.  This is an unusual result because ground states of local and translation invariant Hamiltonians are not expected to be so highly entangled.  
Our approach is complementary to recent constructions that also achieve volume-law entanglement~\cite{klich3,klich4}.

\begin{figure}
    \centering
    \includegraphics[width=1\columnwidth]{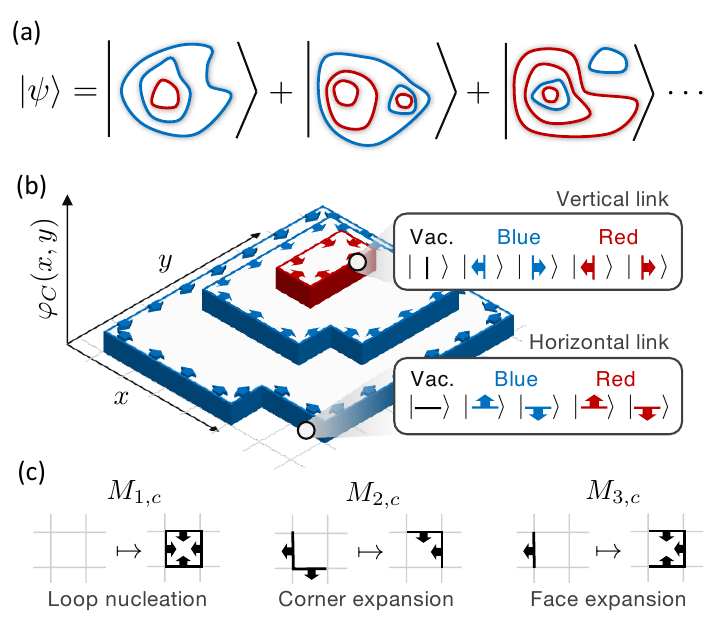}
    \caption{(a) The ground state wavefunction is a superposition of all possible colored, non-crossing, framed loop configurations.
    (b) Loops are formed by local degrees of freedom and can be viewed as contour lines for a 3D landscape, defining the height field $\varphi_C(x,y)$.
    (c) The rules used to construct the kinetic term of the Hamiltonian.  Black lines and arrows can be either red or blue.}
    \label{fig:motzkinfig1}
\end{figure}

\emph{Biased colored loop model} --- We first provide an intuitive picture of the ground state wavefunction of our first model, which possesses anomalous TEE. 
The wavefunction is given as a weighted superposition of all possible configurations of mutually non-crossing, colored loops (Fig.~\ref{fig:motzkinfig1}(a)).
Each loop is formed by chaining together local loop segments equipped with a ``framing'' or ``orientation'' that
allows one to locally  distinguish the interior and exterior of the loop (Fig.~\ref{fig:motzkinfig1}(b)).  
A given configuration $C$ of framed colored loops can be viewed as the contour lines for a 3D landscape, naturally giving rise to a height field $\varphi_C(x,y)$, defined as the number of loops enclosing a spatial point $(x,y)$.
In our model, the quantum amplitude of a loop configuration is determined by the volume of its representative landscape, leading to the explicit wavefunction
\begin{equation}\label{eq:gs}
\ket{\psi(t)} = \frac{1}{\sqrt{Z}}\sum_{C} t^{\sum_{x,y} \varphi_C(x,y)} \ket{C},
\end{equation}
where $C$ enumerates over configurations of loops, $t > 1$ is a tunable parameter, and $Z$ is the normalization factor. 
This state is ``biased'' in the sense that typical configurations in the superposition  favor landscapes with a single large ``mountain'' of concentric loops.

\begin{figure}
    \centering
    \includegraphics[width=1\columnwidth]{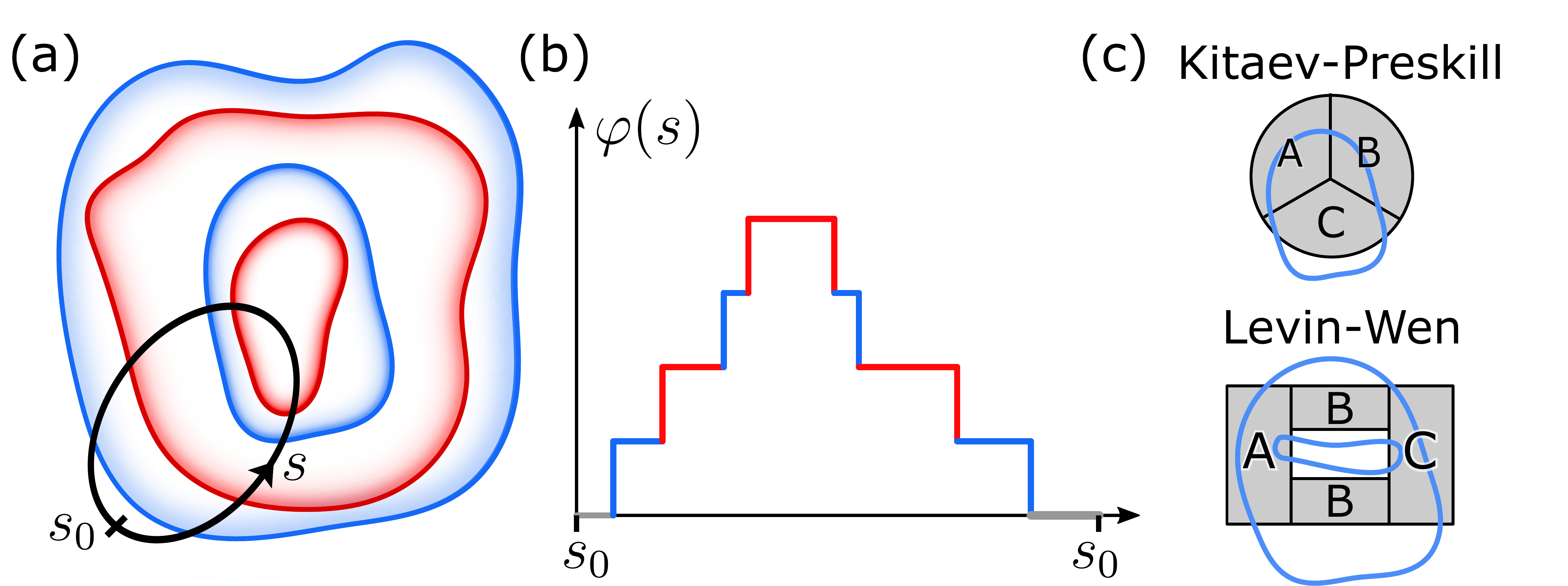}
    \caption{Panel (a) shows a typical configuration in the ground state of the biased colored loop model.  The black path in (a) traces a sequence of height fields indicated in (b), which can be associated to a set of matched colored parentheses.  Computing the TEE requires counting the expected number of loops crossing regions $ABCD$ in the Kitaev-Preskill partition (c) and $ACD$ in the Levin-Wen partition (d), where $D = (ABC)^c$.}
    \label{fig:motzkinfig2}
\end{figure}

Intuitively, this state must have unusual entanglement properties.  This is because typical configurations involve many long loops which carry long range correlations due to their color and non-crossing properties.
More explicitly, imagine a closed path on the 2D lattice, for example in Fig.~\ref{fig:motzkinfig2}(a).
When one traverses the path, the exact number, color, and order of loops which were entered has to match with those of loops which were exited, analogous to the Hilbert space constraint of the Motzkin chain~\cite{movassagh} which features configurations of matched colored parentheses (Fig.~\ref{fig:motzkinfig2}(b)).
This leads to a large amount of non-local correlations.

Now we construct a parent Hamiltonian for $\ket{\psi(t)}$ that is frustration free, geometrically local, and translationally invariant.
This is done in three steps.
First, we identify a local Hilbert space, with loop and framing degrees of freedom.
Second, we construct a diagonal Hamiltonian $H_\textrm{con}$ that 
% enforces a constraint 
restricts the ground state to a subspace spanned by all possible non-crossing loop configurations. 
Finally, we add an off-diagonal kinetic term $H_\textrm{kin}$ designed such that $H= H_\textrm{con} + H_\textrm{kin}$ admits $\ket{\psi(t)}$ as its unique ground state.

Concretely, we consider an $L\times L$ square lattice with open boundary conditions.  The local Hilbert space $\mch$ is defined on the links of the lattice, with an orthonormal basis denoted by $\ket{|}$, $\ket{\textcolor{blue}{\leftframe_{\,b}}}$, $\ket{\textcolor{blue}{\rightframe_{\,b}}}$, $\ket{\textcolor{red}{\leftframe_{\,r}}}$, $\ket{\textcolor{red}{\rightframe_{\,r}}}$ for vertical links and
%$\ket{\text{---}}$ $\ket{\textcolor{blue}{\upframe_{\,b}}}$, $\ket{\textcolor{blue}{\downframe_b}}$, $\ket{\textcolor{red}{\upframe_{\,r}}}$, $\ket{\textcolor{red}{\downframe_r}}$ 
similarly for horizontal links (Fig~\ref{fig:motzkinfig1}(b)).
The subscript denotes the color, b(lue) or r(ed), which can be easily generalized to more than two. The height fields are defined on the sites of the dual lattice.  To achieve a collection of colored non-intersecting loops on this lattice, we choose $H_{\text{con}}$ to energetically favor configurations having either zero or two identically-colored bonds ending on any given site by imposing energy penalty to all other configurations. In addition, we energetically disallows a small loop whose framing points in the wrong direction.

What remains is the construction of $H_{\text{kin}}$. 
Each term in $H_{\text{kin}}$ is associated with an ``update rule'' that modifies the loop configuration.
There are three groups of rules (Fig~\ref{fig:motzkinfig1}(c)).
The first rule $M_{1,c}$ with $c\in \{b,r\}$ nucleates a loop of color $c$ from a (local) vacuum configuration, with all framing vectors pointing inwards. Rules $M_{2,c,i}$ and $M_{3,c,i}$ deform an existing loop while preserving the framing; the subscript $i$ is introduced to describe the similar move in four different orientations.
These rules are conditioned on the new configuration not violating any terms in $H_{\text{con}}$, which can be checked locally. 
From these rules, we define 
\begin{align}
    H_{\text{kin}} = &\sum_{\text{all rules } M}(t \ket{C} - \ket{M(C)})(t \bra{C} - \bra{M(C)})
\end{align}
where $M(C)$ is the configuration obtained after applying $M$ to $C$.
The factors of $t$ control the ratio of amplitudes associated with $C$ and $M(C)$.

We claim that $H=H_{\rm kin}+H_{\rm con}$ is frustration-free and has the desired ground state in Eq.~\eqref{eq:gs}. Clearly $H$ is positive semidefinite, and so any zero energy eigenstate is a ground state.  Such a state $\ket{\psi}$ must be annihilated by each term in $H_{\rm kin}$, and hence must satisfy $t \braket{C}{\psi} = \braket{M(C)}{\psi}$ for all $M$. One can check that these are satisfied by Eq.~\eqref{eq:gs}: all rules increase the volume of the loop configuration by 1, and hence the amplitude of $\ket{M(C)}$ must be $t$ times the amplitude of $\ket{C}$.
We note that the framing of loops plays a crucial role here; it allows to \emph{locally} distinguish if a move increases or decreases the volume.

To show the uniqueness, we invoke the ergodicity of the rules; any two configurations satisfying the constraints in $H_{\rm con}$ can be connected by a sequence of rules. This holds because any configurations can be deformed to the vacuum state, from which one can get to any other collection of loops.
The ergodicity and the condition $t \braket{C}{\psi} = \braket{M(C)}{\psi}$ imply that Eq.~\eqref{eq:gs} must be the unique ground state.

\emph{Entanglement scaling} --- We claim that the above ground state exhibits area law entanglement scaling. 
Consider a vertical cut dividing the lattice into region $A$ and its complement $A^c$.
We use the value of the height fields along $\partial A$ 
to construct the Schmidt decomposition:
\begin{equation}\label{eq:schmidt}
\ket{\psi} = \sum_{\varphi_b}t^{\sum\nolimits_{\mathbf{x} \in \partial A} \varphi_b(\mathbf{x})}\sqrt{\frac{Z_{\varphi_b}(A) Z_{\varphi_b}(A^c)}{Z}} \ket{\psi_{\varphi_b}(A)} \otimes \ket{\psi_{\varphi_b}(A^c)}
\end{equation}
where 
\begin{equation}
    \ket{\psi_{\varphi_b} (A)} = \frac{1}{\sqrt{Z_{\varphi_b}(A)}} 
    \sum_{C: \varphi_C (\partial A)=\varphi_b} t^{\sum_{\mathbf{x}\in A} \varphi_C(\mathbf{x})} \ket{C}
\end{equation}
is the wavefunction restricted to the subsystem $A$ with a constraint on the  height field at the boundary $\varphi_b$ and a normalization constant $Z_{\varphi_b} (A)$.
We note the orthogonality relations
\begin{equation}
\braket{\psi_{\varphi_b}(A)}{\psi_{\varphi'_b}(A)} = \delta_{\varphi_b, \varphi'_b}
\end{equation}
and similarly for $A^c$, therefore establishing that Eq.~\eqref{eq:schmidt} is indeed the Schmidt decomposition for $\ket{\psi}$.  The number of non-zero Schmidt coefficients is the number of distinct height field assignments to $\partial A$, which is upper bounded by $({\rm dim}\, \mch)^{|\partial A|}$; therefore, the entanglement entropy is upper bounded by $S(\rho_A) \leq |\partial A| \log {\rm dim}\, \mch$, which implies area law entanglement.  The exact
% formal expression for the 
entanglement entropy is $S(\rho_A) = -\sum_{\varphi_b} p(\varphi_b) \log p(\varphi_b)$, where
\begin{equation}
\label{pphi}
p(\varphi_b) = t^{2\sum_{\mathbf{x} \in \partial A} \varphi_b(\mathbf{x})}\frac{Z_{\varphi_b}(A) Z_{\varphi_b}(A^c)}{Z}.
\end{equation}
Intuitively, $p(\varphi_b)$ is the probability of measuring a boundary configuration $\varphi_b$.  Based on typical configurations in the biased colored loop model, the most probable configuration of $\varphi_b$ will increase up to the middle of the cut and subsequently decrease.  
The multiplicity of this 
configuration is roughly $|c|^{O(L)}$ with $|c|=2$ number of colors, and therefore the entropy of a typical configuration yields an area law term. We emphasize that this is not a conventional contribution to the area law, as it comes from long range correlations across the entanglement cut.  Thus, on top of this exotic contribution, we expect an additional conventional area law contribution originating from local fluctuations of loops or equivalently height fields.  We confirm area law scaling for an arbitrary region $A$ in the SM~\cite{si}.

\emph{Topological entanglement entropy} ---   
There are two popular schemes for computing the TEE, one due to Kitaev and Preskill (KP) and the other due to Levin and Wen (LW).  In the KP scheme \cite{KitaevPreskill}, a circular region is split into three wedges $A$, $B$, $C$, and one computes the \emph{quantum interaction information} $I(A:B:C) = S(A) + S(B) + S(C) - S(AB) - S(BC) - S(AC) + S(ABC)$.  For a topologically ordered system with $S(A) = \alpha |\partial A| - \gamma$, one finds $I(A:B:C)=-\gamma$. In the LW prescription \cite{LevinWen}, the geometry involves sandwiching two disconnected regions $B_1 \cup B_2 \triangleq B$ between regions $A$ and $C$; the TEE is given by the conditional mutual information $I(A;C|B)$, which for 
the same entanglement scaling
gives $\gamma$. These geometries are designed to eliminate the contribution of area law terms, assuming the area law arises from {\it local} entanglement. The biased colored loop model presents a counterexample where cancellation does not occur.

In order to demonstrate this, we introduce a biased \emph{colorless} loop model and contrast its TEE scaling against the biased colored loop model.  Specifically, we define a map $Q:\mch \rightarrow \mch_{\text{uncol}} $ into an `uncolored' Hilbert space $\mch_{\text{uncol}}$. $Q$ acts on loops in $\mch$ (labelled by $L$) as $Q(\ket{\textcolor{blue}{L_b}}) = Q(\ket{\textcolor{red}{L_r}}) = 2^{-1/2}\ket{L}$ (for $|c|$ colors, $2^{-1/2}$ is replaced with $|c|^{-1/2}$, ensuring the normalization of $Q(\ket{\psi})$). $Q(\ket{\psi})$ can be argued to be the ground state of a biased colorless loop model, expected to exhibit conventional area law entanglement scaling~\cite{si}.

We now state our main result concerning the difference between the TEE of $\ket\psi$ and $Q(\ket{\psi})$: 
\begin{theorem}
For the biased colored loop model with $|c|$ possible colors, the quantum interaction information $I(A:B:C)$ for the KP geometry satisfies
\begin{equation}
I_{\psi}(A:B:C) - I_{Q(\psi)}(A:B:C) = \langle \mathcal{L}_{ABCD} \rangle \log |c|
\end{equation}
where $D = (ABC)^c$ and $\langle \mathcal{L}_{X_1X_2\dots} \rangle$ denotes the expected number of loops that extend into all of regions $X_i$ and not more.
\end{theorem}
\begin{theorem} 
The conditional mutual information $I(A;C|B)$ for the LW geometry satisfies
\begin{equation}
I_{\psi}(A;C|B) - I_{Q(\psi)}(A;C|B) = \langle \mathcal{L}_{ACD} \rangle \log |c|.
\end{equation}
\end{theorem}
These theorems imply anomalous TEE for the biased colored loop model, and their detailed proofs are presented in SM~\cite{si}.  It is important to note that the nonzero value for the TEE in our model does not imply nontrivial topological degeneracy.  One should instead treat the TEE as a probe of non-local entanglement, which the biased colored loop model has an unusually large amount of, despite exhibiting area law scaling. In particular, since typical configurations consist of a single mountain of loops, loops are expected to have $O(L)$ length on average. For the KP prescription, an example of a loop crossing regions $A,B,C,D$ is shown in the top panel of Fig.~\ref{fig:motzkinfig2}(c).  We can consider a generic region $ABC$, which for simplicity we model as an ellipse with major axis $\ell$ and minor axis $\ell(1-\epsilon)$.  Then, for typical configurations, the number of loops crossing into all four regions is $O(\epsilon \ell)$.  Therefore, at least one of $I_{\psi}(A:B:C)$ and $I_{Q(\psi)}(A:B:C)$ must have anomalous TEE, i.e. a TEE which is proportional to the size of the region under consideration.  Clearly it must be $\ket\psi$ which is anomalous, because for any 1D closed loop of length $\ell$ the biased colored loop model mimics the colored Motzkin chain with an added volume law deformation: the entanglement scaling of this loop is therefore $O(\ell)$ and yields a nonlocal contribution to the area law term.  In contrast, for the uncolored Motzkin chain, the entanglement scaling is $O(1)$ \cite{klich}, so $Q(\psi)$ cannot have an anomalous correction.  Using the LW prescription, an example of loops intersecting regions $A$, $C$, and $D$ is shown in the bottom panel of Fig.~\ref{fig:motzkinfig2}(c).  
By positivity of quantum conditional mutual information, it immediately follows that $I_{\psi}(A;C|B) \geq O(\ell)$ for such geometries.

Another unusual property relates to the relative signs of the KP and LW formulae for the TEE.  For these quantities, it is expected that $
    -I_{\psi}(A:B:C) = I_{\psi}(A;C|B) = \gamma$
in a gapped system, and in general that both quantities have opposite signs (this has been shown in holographic theories~\cite{hayden}).  In the biased colored loop model, the anomalous TEE in both prescriptions is positive -- for the LW prescription this is not surprising as the quantum conditional mutual information is always positive, but to our knowledge, for KP this appears to be the first example of a positive value~\cite{ikim}.  It is an interesting open problem whether a positive KP TEE can exist in a gapped system.

There is a related concept worth noting called \emph{spurious} TEE, where certain gapped systems in a trivial phase exhibit nontrivial TEE~\cite{williamson2019spurious, zou2016spurious, kato2020toy, kim2023universal}.  However, this correction is a constant (rather than an area law) and is believed to be due to the existence of hidden string order made possible in models with subsystem symmetries, qualitatively different from the nonlocal order in the present model.  While the Hamiltonian of our model is presumably somewhat fine-tuned, the anomalous TEE can still detect unusual quantum phase transitions, such as the transition between the short loop phase at $t < 1$ (where the TEE is zero) and the long loop phase at $t > 1$ (where it exhibits area law scaling).  

\emph{Decorated biased loop model} ---   
We now show how to extend this construction to produce a ground state with {\it volume law} entanglement.   
We refer to the resulting model as the \emph{decorated biased loop model}.  

\begin{figure}
    \centering
    \includegraphics[width=1\columnwidth]{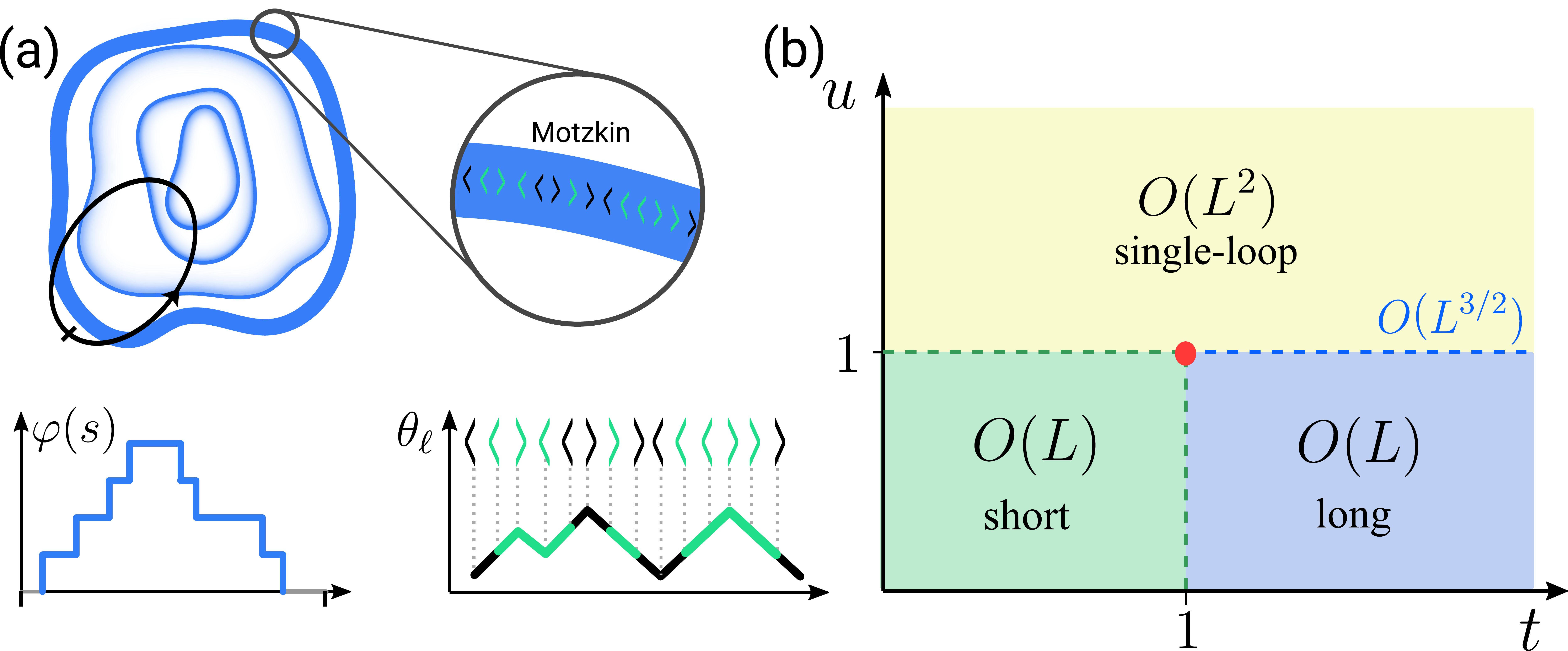}
    \caption{Panel (a) shows a schematic representation of the ground state of the decorated biased loop model, where loops are decorated by a Motzkin chain.  Two different height fields $\varphi$ and $\theta_{\ell}$ are also depicted.  Panel (b) shows a conjectured phase diagram and entanglement scalings for the decorated biased loop model when the number of colors of parentheses is at least 2.  The single color case is discussed in the text.}
    \label{fig:motzkinfig3}
\end{figure}

We start by changing the local Hilbert space of the uncolored ($|c|=1$) model to $
\mathcal{H} = \mathcal{H}_{\text{vac}} \oplus (\mathcal{H}_{\text{arr}} \otimes \mathcal{H}_{\text{mot}})$,
where (fixing our attention to horizontal links) $\mathcal{H}_{\text{vac}}$ is spanned by $\ket{-}$ and $\mathcal{H}_{\text{arr}}$ by $\{\ket{\upframe},\ket{\downframe}\} $ as in the case of our biased colorless loop model.  $\mathcal{H}_{\text{mot}}$ is a new Hilbert space spanned by $\ket{\langle_d}$, $\ket{\rangle_d}$, and $\ket{0}$, where ``$\langle_d$'' and ``$\rangle_d$'' are used to denote left and right parenthesis symbols with $d$ labelling a set of more than one color.  On each loop of the biased uncolored loop model, we then allow the degrees of freedom in $\mathcal{H}_{\text{mot}}$ to form a Motzkin chain with periodic boundary conditions~\cite{movassagh, klich}, i.e. the weighted superposition of all configurations with matched parentheses (Fig.~\ref{fig:motzkinfig3}(a)).  The Motzkin constraint (i.e. enforcing matched parentheses) on open boundary conditions is known to admit a mapping to a height representation denoted by $\theta$, such that $\theta_i$ corresponds to the difference in the number of open and closed parentheses in the interval $[0,i]$.  With periodic boundary conditions, $i = 0$ corresponds to the location where the Motzkin constraint is satisfied if the chain is cut there. Given a loop $\ell$, the the degrees of freedom in $\mathcal{H}_\text{mot}$ along the loop form the state 
\begin{equation}
\ket{\psi_{\ell}(u)} = \frac{1}{\sqrt{Z_{\ell}(u)}} \sum_{C_{\ell}} u^{\sum_i \theta_{\ell}(C_{\ell})} \ket{C_{\ell}},
\end{equation}
where $C_{\ell}$ is the set of configurations satisfying the Motzkin constraint, and the parameter $u$ plays the role of the volume deformation parameter in the Motzkin chain.
Any loop configuration can be mapped onto a set of height fields $\varphi_{x,y}$, as in the biased colored loop model.   
It is then possible to construct a Hamiltonian \cite{si} with ground state
\begin{equation}
\ket{\psi(t, u)} = \frac{1}{\sqrt{Z}} \sum_{C} t^{\sum_{x,y} \varphi_{x,y}(C)} \bigotimes_{\ell \in C} \sqrt{Z_{\ell}(u)}\ket{\psi_{\ell}(u)}.
\end{equation}
i.e. a superposition of biased loops where each loop is decorated with a Motzkin chain (Fig.~\ref{fig:motzkinfig3}(a)).   

 To study the entanglement properties of this state, we note that for $u > 1$ the typical configuration of loops in the ground state is dominated by a single loop of length $O(L^2)$. This is because longer loops are favored due to entropic forces associated with additional degrees of freedom in $\mathcal{H}_\text{mot}$. To compute the bipartite entanglement entropy of this configuration, we must compute a bipartite entanglement entropy associated with cutting the area-weighted Motzkin chain decorating this loop; we expect this to contribute $O(L^2)$ to the entanglement entropy, therefore indicating volume law scaling.  In the SM \cite{si}, we provide a non-rigorous argument for volume law scaling.

Next, we explore the phase diagram obtained by varying $u$ and $t$ (Fig.~\ref{fig:motzkinfig3}(b)).  When $u,t < 1$, loops are small and the Motzkin chains decorating the loops have little entanglement, yielding a conventional area law phase.  When $u < 1$ and $t > 1$, the loops still have little entanglement but are long, still yielding an area law phase.  Both area law phases are distinct since $t = 1$, $u<1$ corresponds to a transition line where the subleading contribution to the entanglement scaling is logarithmic.  The region where $u > 1$ has volume-law scaling, regardless of the value of $t$. Finally, when $u = 1$ and $t > 1$, loops of length $O(L)$ will have bipartite entanglement entropy $O(\sqrt{L})$.  As typical configurations are dominated by a single mountain of $O(L)$ such loops, we expect $S \sim O(L^{3/2})$ (we have argued that this is a reasonable lower bound in \cite{si}). 
What happens at the multicritical point $u = t = 1$ and along the transition line at $t < 1, u = 1$ are questions we leave to future work.

When there is only a single color of parenthesis, the $u>1$ portion of the phase diagram is dominated by a single $O(L^2)$ length loop, however 
the precise entanglement scaling in this phase is unclear. 
By an analogous argument to the $|d| > 1$ case, the transition line $u = 1$ and $t > 1$ has $S \sim O(L \log L)$, reminiscent of models with Fermi surfaces.   
By replacing the decorating Motzkin chain by any $1+1$D CFT, one expects to obtain a similar entanglement scaling.

\emph{Discussion} --- Thus far, we have not discussed loop models whose main fluctuations come from loop merging/splitting processes, paradigmatic examples being the fully packed loop models.  In the SM \cite{si}, we construct an example of a decorated fully-packed loop model which we believe to have high entanglement scaling; it would be interesting to study this model in further depth.

More generally, we believe that our constructions can be easily extended to other kinds of loop models~\cite{freedman2004class} like string net models~\cite{stringnet}, quantum dimer models~\cite{RK}, or quantum vertex models~\cite{ardonne2004topological, castelnovo2005quantum, Balasubramanian, Balasubramanian2}.  Generalizing to 3D should also be possible: in this case, one has a choice of whether the Hilbert space constraint involves dimer or plaquette variables.  If the former, one may decorate loops of dimers with 1D Motzkin chains, and if the latter, one may decorate closed surfaces of plaquettes with the 2D models introduced in this paper.  There are also alternative ways of constructing loop models with long loop phases in 3D~\cite{nahum1, nahum2}, and it would be interesting to study decorated analogues of these models. Our construction can also be modified by decorating loops with other 1D spin chains that exhibit high entanglement, such as ones in Refs.~\cite{korepin2, klich2, caha, padmanabhan2019quantum}.  It would also be interesting to consider dynamical properties of our models; robust ergodicity breaking phases were recently constructed in a model with a mapping onto loop configurations (see Ref.~\cite{stephen2022ergodicity} and recently Ref.~\cite{stahl}), and it is natural to wonder if our model can be modified to exhibit similar properties.

\emph{Acknowledgements} --- We thank Aram Harrow, Isaac Kim, Sarang Gopalakrishnan, Ruben Verresen, Dan Ranard, Israel Klich, and Zhao Zhang for useful comments and helpful discussions.  S.B. was supported by the National Science Foundation Graduate Research Fellowship under Grant No. 1745302.  E.L. was supported by the Hertz Foundation Fellowship.

\bibliography{cit}

\onecolumngrid
	\begin{center}
		
        \bigskip 
        \bigskip 
\textbf{\large Supplementary material for:\\ 2D Hamiltonians with exotic bipartite and topological entanglement}\\[.3cm]		
		Shankar Balasubramanian$^{1,2}$, Ethan Lake$^{2}$, Soonwon Choi$^{1,2}$ \\[.1cm]
		{\itshape ${}^1$Center for Theoretical Physics, Massachusetts Institute of Technology, Cambridge, MA 02139, USA\\
			${}^2$Department of Physics, Massachusetts Institute of Technology, Cambridge, MA 02139, USA \\
		(Dated: \today)}\\[1cm]
	\end{center}
	
	\setcounter{equation}{0}
	\setcounter{figure}{0}
	\setcounter{table}{0}
	\setcounter{page}{1}
	\renewcommand{\theequation}{S\arabic{equation}}
	\renewcommand{\thefigure}{S\arabic{figure}}

	\subsection*{Contents} 
	\begin{itemize}
		\item Section \ref{app:bclham}: construction of the Hamiltonian for the biased colored loop model. 
		\item Section \ref{app:theorems12}: proofs of Theorems 1 and 2 in the main text. 
        \item Section \ref{app:rigorous_tee}: detailed analysis of the TEE.
        \item Section \ref{app:bcl_bconds}: a discussion of boundary conditions in the biased colored loop model. 
        \item Section \ref{app:subsystem}: a discussion of subsystem symmetry breaking as a mechanism for anomalous TEE and an argument that the biased colored loop model is more robust. 
        \item Section \ref{app:ham_decorated_bcl}: construction of the Hamiltonian for the decorated biased colored loop model.
        \item Section \ref{app:volume_law}: a heuristic argument for volume-law entanglement in the ground state of the decorated model for $u>1$. 
        \item Section \ref{app:u1entang}: a heuristic argument for entanglement scaling in the ground state of the decorated model when $u=1$.
        \item Section \ref{app:phase_diagram}: a summary of the phase diagram of the decorated model, along with open questions.
        \item Section \ref{app:single_color_parenthesis}: the decorated model with a single color of parenthesis.
        \item Section \ref{app:decorated_packed_loop}: the fully-packed version of the decorated model. 
	\end{itemize}
	
	%\bs 

\section{Details of the Hamiltonian for the biased colored loop model}\label{app:bclham}

In this section, we present explicitly a frustration-free Hamiltonian whose ground state has the advertised anomalous topological entanglement entropy.  Our model is defined on an $L \times L$ square lattice for simplicity, although it is generalizable to other lattices.

In the main text, we defined the Hamiltonian of biased colored loop model on a direct square lattice.  In this appendix, we will present an explicit construction of the Hamiltonian on the \emph{dual lattice} of the square lattice.
  
Specifically, this dual lattice is formed in the following way.
Starting from the original square lattice, we introduce one ``site'' (or vertex) at the center of every plaquette of the original lattice. Then, all nearest-neighboring site pairs are connected by either vertical horizontal links.
In this way, quantum degrees of freedom, originally placed at the links of the original lattice, are located at the links of the dual lattice. Note that the quantum degrees of freedom associated horizontal (vertical) links in the original lattice are now associated with vertical (horizontal) links in the dual lattice.
This contruction can be generalized for any lattice associated with a planar graph.
In what follows, the lattice refers to the dual lattice unless specified otherwise.

The degrees of freedom are dimer-like variables defined on the links of the lattice. 
The local Hilbert space on the links of the lattice will be denoted as $\mathcal{H} = \text{span}\{\ket{-}, \ket{\leftarrow_c}, \ket{\rightarrow_c}\}$ on  horizontal links and $\mathcal{H} = \text{span}\{\ket{|}, \ket{\uparrow_c}, \ket{\downarrow_c}\}$ on vertical links.
The arrows provide a certain chirality to the dimer variables.  The label $c$ denotes a color assigned to the dimers, and in the analysis that follows we require a number of colors $\geq 2$.
We note that in the lattice described in the main text, $\text{span}\{\ket{-}, \ket{\leftarrow_c}, \ket{\rightarrow_c}\}$ maps to $\text{span}\{\ket{|}, \ket{\leftframe_c}, \ket{\rightframe_c}\}$ and $\text{span}\{\ket{|}, \ket{\uparrow_c}, \ket{\downarrow_c}\}$ maps to $\text{span}\{\ket{\text{---}}, \ket{\upframe_c}, \ket{\downframe_c}\}$.  This identification can be made to map between the direct lattice representation of the Hamiltonian in the main text and dual lattice representation here.

\subsection{Diagonal terms constraining the Hilbert space}

We first impose effective constraints on the many-body Hilbert space.
Since we are aiming to construct a frustration-free Hamiltonian, introducing any non-vanishing energy penalty to a certain set of configurations will suffice to effectively constraining the Hilbert space.
Around each plaquette, we add energy penalizing terms for configurations that are \emph{not} one of the following types:
\begin{equation}\label{eq:allowedconfig1}
\begin{array}{ccccc}
\begin{tikzpicture}[baseline={([yshift=-.5ex]current bounding box.center)},vertex/.style={anchor=base,circle,fill=black!25,minimum size=18pt,inner sep=2pt}]
    \draw (0,0) -- (0.8,0) -- (0.8,0.8) -- (0,0.8) -- (0,0); 
  \end{tikzpicture} & \begin{tikzpicture}[baseline={([yshift=-1.5ex]current bounding box.center)},vertex/.style={anchor=base,circle,fill=black!25,minimum size=18pt,inner sep=2pt}]
    \draw (0,0) -- (0.8,0) -- (0.8,0.8) -- (0,0.8) -- (0,0); 
    \draw[->] (0.8,0.8)--(0.4,0.8) node[above] {$c$};
    \draw[->] (0,0)--(0,0.4) node[left] {$c$};
    \node[right] at (0.8,0.4) {$\text{ }$};
  \end{tikzpicture} & \begin{tikzpicture}[baseline={([yshift=-1.5ex]current bounding box.center)},vertex/.style={anchor=base,circle,fill=black!25,minimum size=18pt,inner sep=2pt}]
    \draw (0,0) -- (0.8,0) -- (0.8,0.8) -- (0,0.8) -- (0,0); 
    \draw[->] (0.8,0)--(0.8,0.4) node[right] {$c$};
    \draw[->] (0,0.8)--(0.4,0.8) node[above] {$c$};
  \end{tikzpicture} & \begin{tikzpicture}[baseline={([yshift=1ex]current bounding box.center)},vertex/.style={anchor=base,circle,fill=black!25,minimum size=18pt,inner sep=2pt}]
    \draw (0,0) -- (0.8,0) -- (0.8,0.8) -- (0,0.8) -- (0,0); 
    \draw[->] (0,0)--(0.4,0) node[below] {$c$};
    \draw[->] (0.8,0.8)--(0.8,0.4) node[right] {$c$};
  \end{tikzpicture} & \begin{tikzpicture}[baseline={([xshift=1ex, yshift=1ex]current bounding box.center)},vertex/.style={anchor=base,circle,fill=black!25,minimum size=18pt,inner sep=2pt}]
    \draw (0,0) -- (0.8,0) -- (0.8,0.8) -- (0,0.8) -- (0,0); 
    \draw[->] (0.8,0)--(0.4,0) node[below] {$c$};
    \draw[->] (0,0.8)--(0,0.4) node[left] {$c$};
  \end{tikzpicture}
  \end{array}
\end{equation}
as well as
\begin{equation}\label{eq:allowedconfig2}
\begin{array}{cccc}
  \begin{tikzpicture}[baseline={([yshift=1.5ex]current bounding box.center)},vertex/.style={anchor=base,circle,fill=black!25,minimum size=18pt,inner sep=2pt}]
    \draw (0,0) -- (0.8,0) -- (0.8,0.8) -- (0,0.8) -- (0,0); 
    \draw[->] (0,0)--(0.4,0) node[below] {$c$};
    \draw[->] (0,0.8)--(0.4,0.8) node[above] {$c$};
  \end{tikzpicture} & \begin{tikzpicture}[baseline={([yshift=1.5ex]current bounding box.center)},vertex/.style={anchor=base,circle,fill=black!25,minimum size=18pt,inner sep=2pt}]
    \draw (0,0) -- (0.8,0) -- (0.8,0.8) -- (0,0.8) -- (0,0); 
    \draw[->] (0.8,0)--(0.8,0.4) node[right] {$c$};
    \draw[->] (0,0)--(0,0.4) node[left] {$c$};
  \end{tikzpicture} & \begin{tikzpicture}[baseline={([yshift=2ex]current bounding box.center)},vertex/.style={anchor=base,circle,fill=black!25,minimum size=18pt,inner sep=2pt}]
    \draw (0,0) -- (0.8,0) -- (0.8,0.8) -- (0,0.8) -- (0,0); 
    \draw[->] (0,0)--(0.4,0) node[below] {$c'$};
    \draw[->] (0.8,0)--(0.8,0.4) node[right] {$c$};
    \draw[->] (0,0.8)--(0.4,0.8) node[above] {$c$};
    \draw[->] (0,0)--(0,0.4) node[left] {$c'$};
  \end{tikzpicture} & \begin{tikzpicture}[baseline={([yshift=2ex]current bounding box.center)},vertex/.style={anchor=base,circle,fill=black!25,minimum size=18pt,inner sep=2pt}]
    \draw (0,0) -- (0.8,0) -- (0.8,0.8) -- (0,0.8) -- (0,0); 
    \draw[->] (0.8,0)--(0.4,0) node[below] {$c'$};
    \draw[->] (0.8,0)--(0.8,0.4) node[right] {$c'$};
    \draw[->] (0.8,0.8)--(0.4,0.8) node[above] {$c$};
    \draw[->] (0,0)--(0,0.4) node[left] {$c$};
  \end{tikzpicture}\\
\end{array}
\end{equation}
as well as all configurations corresponding to reversing all of the arrows in the configurations above.  Note that in the above, $c \neq c'$.  We may either choose to include or exclude the last two configurations above in the constrained Hilbert space-- our final results will not change either way.  \emph{For clarity of presentation in the main text, we worked in the constrained Hilbert space where the last two configurations above were omitted}, so that on the dual lattice these constraints appropriately correspond to configurations of self avoiding loops.

We introduce one more term, which provides a constraint corresponding to energetically disallowing configurations $\ket{\mathcal{V}_{v,c}}$ with all arrows of color $c$ pointing out of vertex $v$.  These configurations are defined around the vertices of the dual lattice and look like
\begin{equation}
   \mathcal{V}_{v,c} = \begin{tikzpicture}[baseline={([yshift=-.5ex]current bounding box.center)},vertex/.style={anchor=base,circle,fill=black!25,minimum size=18pt,inner sep=2pt}]
    \draw (-0.8,0) -- (0.8,0);
    \draw (0,-0.8) -- (0,0.8); 
    \draw[<-] (-0.4,0) -- (0,0);
    \draw[<-] (0,-0.4) -- (0,0);    
    \draw[->] (0,0) -- (0.4,0);
    \draw[->] (0,0) -- (0,0.4);
  \end{tikzpicture}
\end{equation}
As it will become more clear after the next subsection, this term is needed to enforce that the unique ground state cannot admit loops with improper orientation.

We note that imposing energy penalties to disallowed configurations is equivalent to giving energy incentives to allowed configurations. Hence we use these phrases interchangeably.  Calling $\mathscr{C}$ the set of all the allowed configurations in Eqn.~\ref{eq:allowedconfig1} and Eqn.~\ref{eq:allowedconfig2} and labelling plaquettes on the square lattice by $p$, we define the projector
\begin{equation}
H_{\text{con}} = -\sum_{C \in \mathscr{C}, \hspace{0.025cm} p} \ketbra{C_p}{C_p} + \sum_{v,c} \ketbra{\mathcal{V}_{v,c}}{\mathcal{V}_{v,c}}
\end{equation}
with the first term locally imposing the desired Hilbert space constraint and the second term expelling incorrectly oriented loops from the ground state subspace. 

\subsection{Off-diagonal kinetic terms favoring a  weighted superposition state}
Next, we discuss kinetic energy terms which result from quantum fluctuations on this constrained space.  As described in the main text, we first introduce the notion of a {\it process} and then devise a Hamiltonian term associated with each process. 
A process $(C, C')$ corresponds to a pair of local dimer configurations $C$ and $C'$ which have a nonzero transition amplitude $C \leftrightarrow C'$. For dimer configurations which are not locally related, a sequence of processes may provide a path between them.

Given a process $(C, C')$, we associate a Hamiltonian term obtained from the local projection operator
\begin{equation}
\mathcal{P}(\ket{C} - \ket{C'}) = (\ket{C} - \ket{C'})(\bra{C} - \bra{C'}),
\end{equation}
which annihilates the superposition $\ket{C} + \ket{C'}$.  If we sum over all possible local processes, yielding
\begin{equation}
H = \sum_{C, C'} \mathcal{P}(\ket{C} - \ket{C'}),
\end{equation}
this Hamiltonian admits a ground state $\ket{\psi_0} = \sum_{C \sim C_0} \ket{C}$, where $C \sim C_0$ means that configuration $C$ can be obtained from reference configuration $C_0$ through repeatedly applying some sequence of local processes.

In order to fully specify the kinetic terms of the Hamiltonian, it only remains to specify all allowed local processes.  We enumerate all processes below.
\begin{align}\label{eqn:processes}
    M_{1,c} &= \begin{tikzpicture}[baseline={([yshift=-.5ex]current bounding box.center)},vertex/.style={anchor=base,circle,fill=black!25,minimum size=18pt,inner sep=2pt}]
    \draw (-0.8,0) -- (0.8,0);
    \draw (0,-0.8) -- (0,0.8); 
    \draw[dashed] (-0.8,-0.8) -- (0.8,-0.8);
    \node[align=center] at (0.4,0.4) {$p_1$};
    \node[align=center] at (-0.4,0.4) {$p_2$};
    \node[align=center] at (-0.4,-0.4) {$p_3$};
    \node[align=center] at (0.4,-0.4) {$p_4$};
    \draw[dashed] (0.8,0.8) -- (0.8,-0.8);
    \draw[dashed] (-0.8,-0.8) -- (-0.8,0.8);
    \draw[dashed] (-0.8,0.8) -- (0.8,0.8);
  \end{tikzpicture} \Longleftrightarrow \begin{tikzpicture}[baseline={([yshift=-.5ex]current bounding box.center)},vertex/.style={anchor=base,circle,fill=black!25,minimum size=18pt,inner sep=2pt}]
    \draw (-0.8,0) -- (0.8,0);
    \draw (0,-0.8) -- (0,0.8); 
    \draw[->] (-0.8,0) -- (-0.4,0);
    \draw[->] (0,-0.8) -- (0,-0.4);    
    \draw[<-] (0.4,0) -- (0.8,0);
    \draw[<-] (0,0.4) -- (0,0.8);
    \draw[dashed] (-0.8,-0.8) -- (0.8,-0.8);
    \draw[dashed] (0.8,0.8) -- (0.8,-0.8);
    \draw[dashed] (-0.8,-0.8) -- (-0.8,0.8);
    \draw[dashed] (-0.8,0.8) -- (0.8,0.8);
  \end{tikzpicture}\nonumber\\
  M_{2,c,i} &= \begin{tikzpicture}[baseline={([yshift=-.5ex]current bounding box.center)},vertex/.style={anchor=base,circle,fill=black!25,minimum size=18pt,inner sep=2pt}]
    \draw (-0.8,0) -- (0.8,0);
    \draw (0,-0.8) -- (0,0.8); 
    \draw[->] (-0.8,0) -- (-0.4,0);
    \draw[->] (0,-0.8) -- (0,-0.4);    
    %\draw[<-] (0.4,0) -- (0.8,0) node[below];
    %\draw[<-] (0,0.4) -- (0,0.8) node[above right];
    \draw[dashed] (-0.8,-0.8) -- (0.8,-0.8);
    \draw[dashed] (0.8,0.8) -- (0.8,-0.8);
    \draw[dashed] (-0.8,-0.8) -- (-0.8,0.8);
    \draw[dashed] (-0.8,0.8) -- (0.8,0.8);
  \end{tikzpicture} \Longleftrightarrow \begin{tikzpicture}[baseline={([yshift=-.5ex]current bounding box.center)},vertex/.style={anchor=base,circle,fill=black!25,minimum size=18pt,inner sep=2pt}]
    \draw (-0.8,0) -- (0.8,0);
    \draw (0,-0.8) -- (0,0.8); 
    %\draw[->] (-0.8,0) -- (-0.4,0) node[above left];
    %\draw[->] (0,-0.8) -- (0,-0.4) node[left];    
    \draw[->] (0,0) -- (0.4,0);
    \draw[->] (0,0) -- (0,0.4);
    \draw[dashed] (-0.8,-0.8) -- (0.8,-0.8);
    \draw[dashed] (0.8,0.8) -- (0.8,-0.8);
    \draw[dashed] (-0.8,-0.8) -- (-0.8,0.8);
    \draw[dashed] (-0.8,0.8) -- (0.8,0.8);
  \end{tikzpicture} \nonumber\\
  M_{3,c,i} &= \begin{tikzpicture}[baseline={([yshift=-.5ex]current bounding box.center)},vertex/.style={anchor=base,circle,fill=black!25,minimum size=18pt,inner sep=2pt}]
    \draw (-0.8,0) -- (0.8,0);
    \draw (0,-0.8) -- (0,0.8); 
    \draw[->] (-0.8,0) -- (-0.4,0);
    %\draw[->] (0,-0.8) -- (0,-0.4) node[left];    
    %\draw[<-] (0.4,0) -- (0.8,0) node[below];
    %\draw[<-] (0,0.4) -- (0,0.8) node[above right];
    \draw[dashed] (-0.8,-0.8) -- (0.8,-0.8);
    \draw[dashed] (0.8,0.8) -- (0.8,-0.8);
    \draw[dashed] (-0.8,-0.8) -- (-0.8,0.8);
    \draw[dashed] (-0.8,0.8) -- (0.8,0.8);
  \end{tikzpicture} \Longleftrightarrow \begin{tikzpicture}[baseline={([yshift=-.5ex]current bounding box.center)},vertex/.style={anchor=base,circle,fill=black!25,minimum size=18pt,inner sep=2pt}]
    \draw (-0.8,0) -- (0.8,0);
    \draw (0,-0.8) -- (0,0.8); 
    %\draw[->] (-0.8,0) -- (-0.4,0) node[above left];
    \draw[->] (0,0) -- (0,-0.4);    
    \draw[->] (0,0) -- (0.4,0);
    \draw[->] (0,0) -- (0,0.4);
    \draw[dashed] (-0.8,-0.8) -- (0.8,-0.8);
    \draw[dashed] (0.8,0.8) -- (0.8,-0.8);
    \draw[dashed] (-0.8,-0.8) -- (-0.8,0.8);
    \draw[dashed] (-0.8,0.8) -- (0.8,0.8);
  \end{tikzpicture}
\end{align}
where the subscript $c$ designates all of the arrowed links to be of color $c$ and the subscript $i$ indicates that there are 3 other configurations related by $\pi/2$ rotations of the diagrams.  The dashed-line edges denote that there must be a certain constraint on such edges in order for these processes to remain within the constrained Hilbert space.  Call
\begin{equation}
H_{\text{con},p} = -\sum_{C \in \mathscr{C}_p} \ketbra{C}{C}
\end{equation}
i.e. a projector onto valid configurations on plaquette $p$.  Next, we define $V_v = \prod_{p_i \in v} P_{p_i}$, which is a projector enforcing valid configurations on the four plaquettes surrounding site $v$ (labelled $p_1, p_2, p_3, p_4$ above).  If process $M$ (written without subscripts for clarity) corresponds to toggling between configurations $C$ and $C'$, i.e. $C \leftrightarrow C'$, then we denote $M^{(0)} = C$ and $M^{(1)} = C'$.  We may write down the following frustration-free Hamiltonian
\begin{align}
    H_{\text{kin}} = &\sum_{v,c} \mathcal{A}_v(M_{1,c}^{(0)}) \mathcal{A}_v(M_{1,c}^{(1)})\mathcal{P}_v\left(\ket{M_{1,c}^{(0)}} - \ket{M_{1,c}^{(1)}}\right) \nonumber\\ &+ \sum_{v,c,i}\mathcal{A}_v(M_{2,c,i}^{(0)}) \mathcal{A}_v(M_{2,c,i}^{(1)})\mathcal{P}_v\left(\ket{M_{2,c,i}^{(0)}} - \ket{M_{2,c,i}^{(1)}}\right) \nonumber\\ &+ \sum_{v,c,i}\mathcal{A}_v(M_{3,c,i}^{(0)}) \mathcal{A}_v(M_{3,c,i}^{(1)})\mathcal{P}_v\left(\ket{M_{3,c,i}^{(0)}} - \ket{M_{3,c,i}^{(1)}}\right)
\end{align}
where we define
\begin{equation}
\mathcal{A}_v(C) = \mel{C_v}{V_v}{C_v}.
\end{equation}
The product of the two $\mathcal{A}$ factors checks that both $\mathcal{M}^{(0)}$ and $\mathcal{M}^{(1)}$ correspond to configurations satisfying the Hilbert space constraint. That is, if either $C$ or $C'$ is not allowed in the ground state subspace, the corresponding quantum amplitude must vanish, enforcing that the wavefunction always satisfy the Hilbert space constraints.  The Hamiltonian $H = H_{\text{kin}} + H_{\text{con}}$ ($H_{\text{con}}$ being summed over both the plaquette and vertex constraints) therefore has a zero energy ground state corresponding to a superposition of all arrow configurations that are connected to the vacuum configuration via some sequence of the processes tabulated above. The second term in $H_\textrm{con}$ ensures that the equal superposition of all configurations that are not connected to the vacuum configuration must incur a non-zero energy penalty.

\begin{figure}
    \centering
    \includegraphics[scale=0.20]{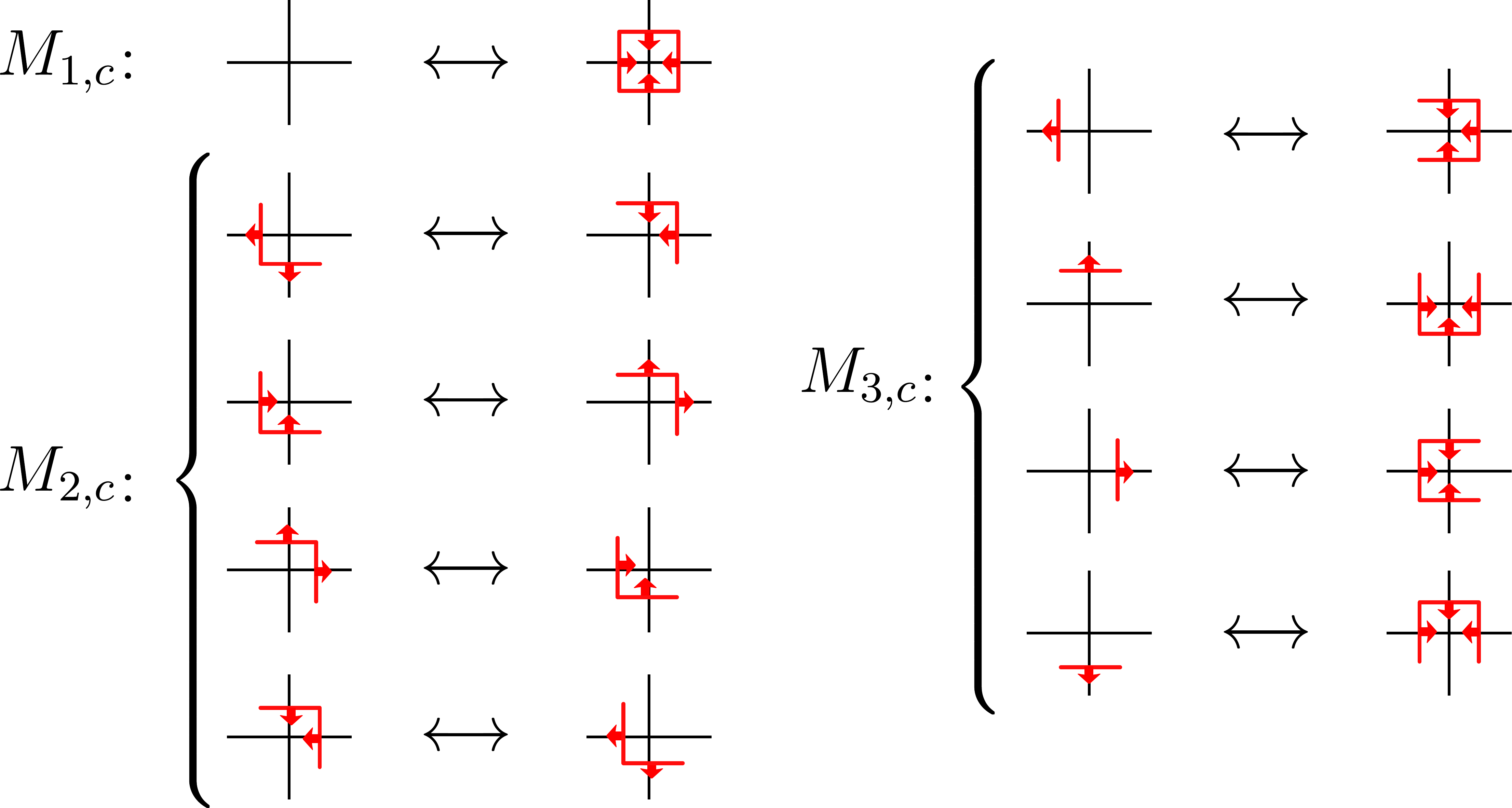}
    \caption{A mapping between the processes $M_{i,c}$ (for $i = 1,2,3$) discussed in the main text, and the framed loop model picture.  The red arrows indicates a framing, pointing in the direction of the interior of the loop.  For $M_{3,c}$, we have 4 more processes corresponding to reversing the arrow direction.  Note that we have an identical set of rules for the arrows being blue.}
    \label{fig:processes}
\end{figure}

We will be working with two equivalent representations of the ground state of the above Hamiltonian.  The first is a mapping to a loop model.  On the direct square lattice formed from the centers of the plaquettes of the dual lattice that we defined the model on, color a link with color $c$ if the link intersects an arrowed link of color $c$ on the dual lattice.  The Hilbert space constraint on the direct lattice corresponds to configurations that are collections of non-crossing closed loops.  The direction of the arrow on the dual lattice is a ``framing'' vector on the direct lattice that locally points in the direction of the interior of the loop.  As shown in Figure~\ref{fig:processes}, on the direct lattice the process $P_1$ corresponds to nucleating a loop from the vacuum, $P_2$ corresponds to moving a corner of a loop, and $P_3$ corresponds to creating a ``bump'' on the loop.  Note that the framing is always preserved under these processes.

The second equivalent representation is a mapping onto a height field representation on the sites of the dual lattice.  Consider a certain configuration $C$, and call $C_m$ the link variable at site $m$.  We choose boundary conditions for $C$ so that the height around the boundaries of the square lattice is 0.  Then, selecting a connected path $\Pi(0), \Pi(1), \cdots, \Pi(L)$ of links from a point on the boundary to $(m,n)$, and denoting the local tangent vector of this path by $\vec{T}(0), \vec{T}(1),\cdots, \vec{T}(L)$, the height $\varphi_{m,n}$ is
\begin{equation}
\varphi_{m,n} = \sum_{i=0}^L \Delta\left(\vec{T}(i), C_{\Pi(i)}\right),
\end{equation}
where $\Delta\left(\vec{T}(i), C_{\Pi(i)}\right)$ is $1$ if $\vec{T}(i)$ points in the same direction as the arrow $C_{\Pi(i)}$, $-1$ if $\vec{T}(i)$ points in the opposite direction as the arrow $C_{\Pi(i)}$, and $0$ if $C_{\Pi(i)}$ is not an arrow.  Therefore, given a height representation, one may uniquely deduce the configuration of arrows.  This height field has an intuitive interpretation in terms of the loop model mapping: the height field $\varphi_{m,n}$ simply counts the number of loops in configuration $C$ that enclose the point $(m,n)$.

As discussed in the main text, we define the volume of a given configuration of dimers $C$ by
\begin{equation}
V(C) = \sum_{x,y} \varphi_{x,y}(C).
\end{equation}
where $\varphi(C)$ maps configuration $C$ into a configuration of height fields $\varphi$.  We now seek a ground state which is a uniform superposition over all configurations connected to the vacuum configuration, with an amplitude weighted by the volume of the configuration:
\begin{equation}
\ket{\psi} \propto \sum_{C \sim C_0} t^{V(C)} \ket{C}.
\end{equation}
As discussed in the main text, this state can be achieved as a ground state of the following local Hamiltonian:
\begin{align}
    H_{\text{kin}}(t) = &\sum_{v,c} \mathcal{A}_v(M_{1,c}^{(0)}) \mathcal{A}_v(M_{1,c}^{(1)})\mathcal{P}_v\left(t \ket{M_{1,c}^{(0)}} - \ket{M_{1,c}^{(1)}}\right) \nonumber\\ &+ \sum_{v,c,i}\mathcal{A}_v(M_{2,c,i}^{(0)}) \mathcal{A}_v(M_{2,c,i}^{(1)})\mathcal{P}_v\left(t \ket{M_{2,c,i}^{(0)}} - \ket{M_{2,c,i}^{(1)}}\right) \nonumber\\ &+ \sum_{v,c,i}\mathcal{A}_v(M_{3,c,i}^{(0)}) \mathcal{A}_v(M_{3,c,i}^{(1)})\mathcal{P}_v\left(t \ket{M_{3,c,i}^{(0)}} - \ket{M_{3,c,i}^{(1)}}\right)
\end{align}
the Hamiltonian $H(t) = H_{\text{con}} + H_{\text{kin}}(t)$ has the desired state as the ground state.

\section{Proofs of theorems 1 and 2}\label{app:theorems12}

We now verify the theorems presented in the main text.  First, note that via an argument from the main text (see around Eq.~\eqref{pphi}), the entanglement entropy of the ground state $|\psi\rangle$ must exhibit area law scaling.  Our claim is that for $t > 1$ and with at least 2 colors of loops, this model exhibits an anomalous topological entanglement entropy (TEE).

\begin{figure}
    \centering
    \includegraphics[scale=0.225]{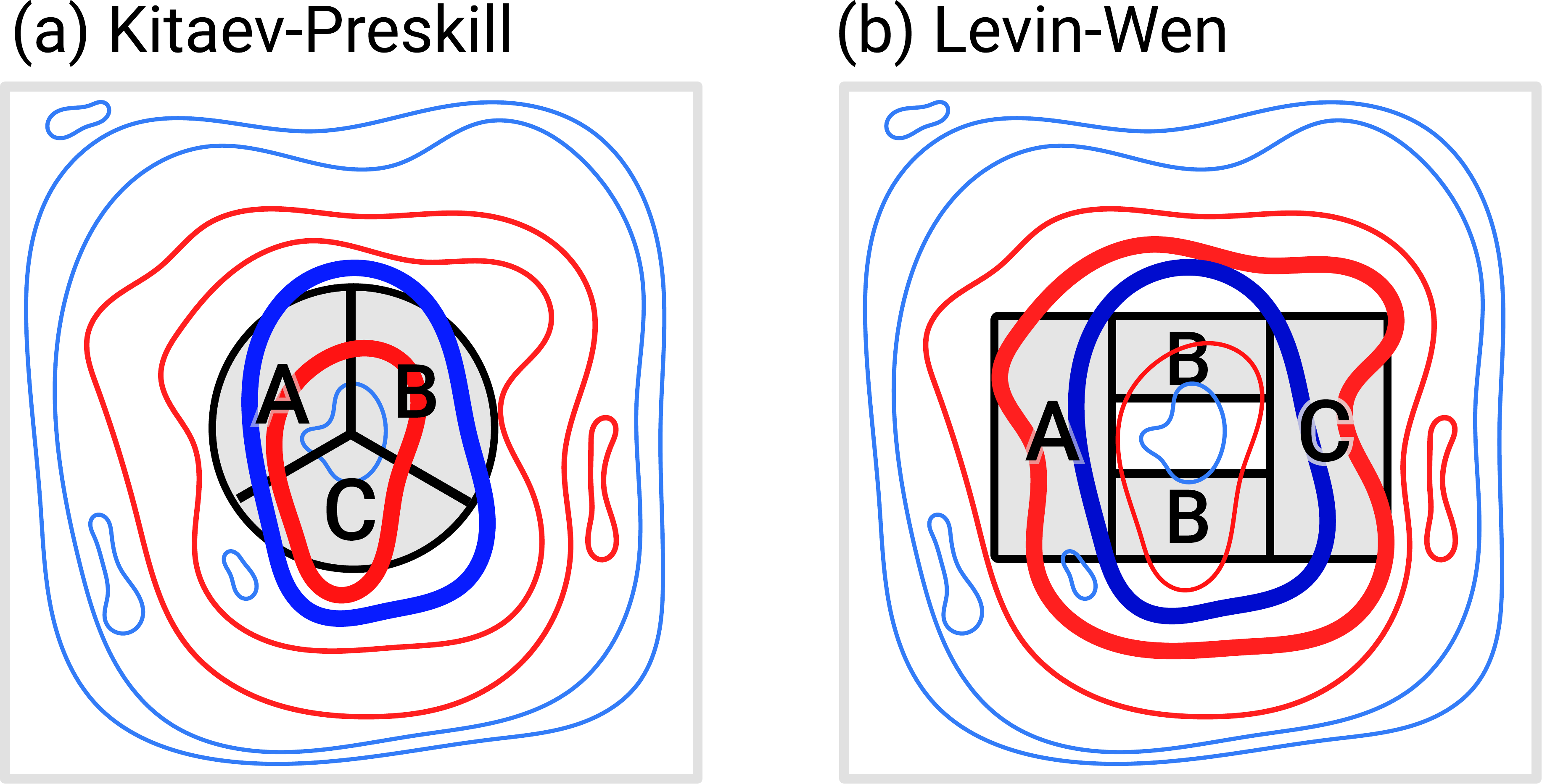}
    \caption{The geometry of the regions used to compute the topological entanglement entropy in the Preskill-Kitaev formulation (panel (a)) and the Levin-Wen formulation (panel (b)) for a particular configuration in the superposition.  The boldened loops indicate the loops which are counted in the computation of the TEE.}
    \label{fig:TEE}
\end{figure}

There are two popular prescriptions for computing the TEE, one based on Ref.~\cite{KitaevPreskill} (Kitaev-Preskill) and the other based on Ref.~\cite{LevinWen} (Levin-Wen).  In the Kitaev-Preskill prescription, one forms a circle with three wedges $A$, $B$, and $C$ and defines
\begin{align}
    \text{TEE}_{\rm KP} = S(A) + S(B) + S(C) - S(AB) - S(BC) - S(AC) + S(ABC),
\end{align}
which is a quantum version of the interaction information $I(A;B) - I(A;B|C)$. The second Levin-Wen type prescription for the TEE is more useful as an information-theoretic quantity because it is equal to a certain conditional quantum mutual information, and is therefore positive.  In this prescription, we sandwich system $B$ between system $A$ and $C$ and split system $B$ into two disconnected components, as shown in Figure~\ref{fig:TEE};  then, the topological entanglement entropy is defined to be
\begin{equation}
    \text{TEE}_{\rm LW} = I(A;C|B) = S(A|B) + S(C|B) - S(AC|B) = S(AB) + S(BC) - S(ABC) - S(B)
\end{equation}
which is positive by virtue of being equivalent to a conditional mutual information.

To compute the TEE for the colored loop model, we first want to compute the entanglement entropy of some subregion $R$ of the system.  Explicitly the boundary $\partial R$ corresponds to a set of connected links on the dual lattice (recall the height fields are defined on the sites of this lattice).  A useful fact is that the ground state wavefunction has a convenient Schmidt decomposition (this Schmidt decomposition was discussed in terms of the height variables in the main text; here we will also discuss the Schmidt decomposition in terms of the loop variables).  In the loop picture, a configuration $C$ has some set of loops $L_n$ which do not intersect with $\partial R$ (the subscript $n$ stands for ``not intersect'') and some set of loops $L_i$ which intersect $\partial R$ ($i$ stands for ``intersect'').  

Only the loops in $L_i$ are necessary to determine the value of $\varphi$ on $\partial R$; therefore, the Schmidt coefficients are labeled only by the intersections of $L_i$ with the entanglement cut.  We label the set of intersection points of the loops with the cut by $I_{\partial R}$.  
Formally, $I_{\partial R}$ is a map between a set of loops and intersection points of these loops on the boundary.  For example, if the set of loops intersecting the boundary is $\{L_1, L_2, \cdots\}$, the intersection point $I_{\partial R}$ is an ordered list of the form $\{L_{1, \rightarrow}, 0, 0, L_{2, \rightarrow},0,0,L_{2, \leftarrow},L_{3, \rightarrow},\cdots\}$ where $0$ indicates no loops intersect at that point, and the subscripts $\leftarrow$ and $\rightarrow$ indicate the framing of the loop.  This data is sufficient to identify a Schmidt decomposition for the ground state.  The relation between $I_{\partial R}$ and the Hilbert space constraint of a colored Motzkin chain living on $\partial R$ is also evident.

For an arbitrary region $R$, the loop intersection points do not by themselves suffice to determine the values of the height fields on $\partial R$.  In particular if $R$ is a disc then $R^c$ is homeomorphic to an annulus (provided the global topology of the system is that of a disc), and homologically nontrivial loops in this annulus induce global shifts of the height field on $\partial R$.  In particular, the height field configuration on $\partial R$ corresponding to a given configuration of loops will be globally shifted by an amount equal to the number of homologically nontrivial loops in $R^c$.  Thus, we must specify both $I_{\partial R}$ and the number of homologically nontrivial loops $L_h$ in $R^c$ when writing a Schmidt decomposition for the ground state wavefunction:
\begin{equation}
    \ket{\psi} = \sum_{I_{\partial R}, L_h} t^{\sum_{\mathbf{x}\in \partial R}|\varphi_{I, L_h}(\mathbf{x})|}\sqrt{\frac{Z_{R^c}(I_{\partial R}, L_h) Z_R(I_{\partial R}, L_h)}{Z}} \ket{\psi_{R^c}(I_{\partial R}, L_h)} \otimes \ket{\psi_R(I_{\partial R}, L_h)}
\end{equation}
where we define the normalized state
\begin{equation}
    \ket{\psi_R(I_{\partial R}, L_h)} = \frac{1}{\sqrt{Z_R(I_{\partial R}, L_h)}}\sum_{\substack{L \in R \\ \left.L\right|_{\partial R} = I_{\partial R}}} t^{\sum_{\mathbf{x}\in R}\varphi_{L,L_h}(\mathbf{x})} \ket{L}
\end{equation}
with the notation $\left.L\right|_{\partial R}$ meaning loop configuration $L$ restricted to the boundary $\partial R$.  Crucially, we may write
\begin{equation}
    \sum_{\mathbf{x}\in R}\varphi_{L, L_h}(\mathbf{x}) = \sum_{\mathbf{x}\in R}\varphi_{L, 0}(\mathbf{x}) + |R| L_h,
\end{equation}
i.e. the effect of the homologically nontrivial loops on the height fields in $R$ is the last term in the expression above.  The partition function (i.e. the normalization of the wavefunction) can also be written as
\begin{equation}
    Z_R(I_{\partial R}, L_h) = t^{2 |R| L_h} Z_R(I_{\partial R}, 0) = t^{2 |R| L_h} \sum_{\substack{L \in R \\ L\left.\right|_{\partial R} = I_{\partial R}}} t^{2 \sum_{\mathbf{x} \in R} \varphi_{L,0}(\mathbf{x})}
\end{equation}
and therefore we may in fact write $\ket{\psi_R(I_{\partial R}, L_h)} = \ket{\psi_R(I_{\partial R}, 0)}$ (as the extra factors of $t^{|R| L_h}$ coming from the amplitudes of the loop configuration cancel with the additional contribution from the partition function).  The states
\begin{equation}
    \ket{\psi_{R^c}(I_{\partial R}, L_h)} = \frac{1}{\sqrt{Z_{R^c}(I_{\partial R}, L_h)}}\sum_{\substack{L \in R \\ \left.L\right|_{\partial R} = I_{\partial R}}} t^{\sum_{\mathbf{x}\in R^c}\varphi_{L}(\mathbf{x})} \ket{L}
\end{equation}
exhibit the orthogonality relations
\begin{equation}
    \braket{\psi_{R^c}(I_{\partial R}, L_h)}{\psi_{R^c}(I'_{\partial R}, L'_h)} = \delta_{I, I'} \delta_{L_h, L'_h}.
\end{equation}
Therefore, the partial trace over $R^c$ can be readily computed:
\begin{equation}
\rho_R = \sum_{I_{\partial R}, L_h} t^{2\sum_{\mathbf{x}\in \partial R}|\varphi_{I, L_h}(\mathbf{x})|} \frac{Z_{R^c}(I_{\partial R}, L_h) Z_R(I_{\partial R}, L_h)}{Z}\ketbra{\psi_{R}(I_{\partial R},0)}{\psi_{R}(I_{\partial R},0)}
\end{equation}

To compute the entropy, we will find it helpful to define $\overline{I}_{\partial R}$ to indicate a set of intersected loops with their colors ignored.  We also abuse notation and interchangeably write $\overline{I}$ instead of $\overline{I}_{\partial R}$, as well as $I$ instead of $I_{\partial R}$.  With this notation, the entanglement entropy is (slightly abusing notation by letting $c$ denote the number of colors):
\begin{equation}
    -\text{Tr}(\rho_R \log \rho_R) = -\sum_{\overline{I}_{\partial R}} c^{L(\overline{I})} \beta_R(\overline{I}_{\partial R}) \log \beta_R(\overline{I}_{\partial R})
\end{equation}
where $L(\overline{I})$ counts the number of loops associated to $\overline{I}$, and
\begin{equation}
   \beta_R(\overline{I}_{\partial R}) = \sum_{L_h} t^{2\sum_{\mathbf{x}\in \partial R}|\varphi_{\overline{I}, L_h}(\mathbf{x})|} \frac{Z_{R^c}(\overline{I}_{\partial R}, L_h) Z_R(\overline{I}_{\partial R}, L_h)}{Z}.
\end{equation}
Defining $p_R(\overline{I}_{\partial R}) = c^{L(\overline{I})} \beta_R(\overline{I}_{\partial R})$, one can verify that $\sum_{\overline{I}_{\partial R}} p_R(\overline{I}_{\partial R}) = 1$ (i.e. $p_R$ is a probability distribution over the colorless $\overline{I}_{\partial R}$)  and consequently
\begin{align}
\label{SrhoR}
    S(\rho_R) = -\sum_{\overline{I}_{\partial R}} p_R(\overline{I}_{\partial R}) \log p_R(\overline{I}_{\partial R}) + \log c \sum_{\overline{I}_{\partial R}} L(\overline{I})\, p_R(\overline{I}_{\partial R}) \triangleq H(p_R) + \langle \ell \rangle \log c
\end{align}
where the notation $\langle \ell \rangle$ denotes the expected number of loops that intersect $\partial R$.  

Recall the ``uncoloring'' map $Q : \mathcal{H} \rightarrow \mathcal{H}_{\rm uncol}$ introduced in the main text, with $Q$ such that the state $Q(\rho_R)$ is associated with superposition of {\it uncolored} framed loops (a configuration of such loops is denoted by $\overline{L}$), where the loops are weighted by both the volume and a factor of $\sqrt{c}^{|\overline{L}|}$:
\begin{equation}
    \ket{Q(\psi)} \propto \sum_{\overline{L}} \sqrt{c}^{|\overline{L}|} t^{V(\overline{L})} \ket{\overline{L}}.
\end{equation}

We claim that the entropy of subregion $R$ for the ``uncolored'' state $Q(\rho_R)$
is given by $S(Q(\rho_R)) = -\sum_{\overline{I}_{\partial R}} p_R(\overline{I}_{\partial R}) \log p_R(\overline{I}_{\partial R}) = H(p_R)$, thus allowing us to complete the evaluation of $S(\rho_R)$.  To show this, we can similarly write this state in terms of a Schmidt decomposition
\begin{equation}
    \ket{Q(\psi)} = \sum_{I_{\partial R}, L_h} t^{\sum_{\mathbf{x}\in \partial R}|\varphi_{I, L_h}(\mathbf{x})|} \sqrt{c}^{L(I)}\sqrt{\frac{\overline{Z}_{R^c}(I_{\partial R}, L_h) \overline{Z}_R(I_{\partial R}, L_h)}{\overline{Z}}} \ket{Q(\psi_{R^c}(I_{\partial R}, L_h))} \otimes \ket{Q(\psi_R(I_{\partial R}, L_h))}
\end{equation}
where the bar in the partition function means
\begin{equation}
    \overline{Z} = \sum_{\overline{L}} c^{|\overline{L}|} t^{2 \sum_{\mathbf{x}}\varphi_{\overline{L}}(\mathbf{x})},
\end{equation}
(which is actually equal to $Z$), and
\begin{equation}
    \overline{Z}_R(\overline{I}_{\partial R}, L_h) = t^{2 |R| L_h} \sum_{\substack{\overline{L} \in R \\ \left.\overline{L}\right|_{\partial R} = \overline{I}_{\partial R}}} c^{|\overline{L}|} t^{2\sum_{\mathbf{x}\in R}\varphi_{\overline{L},0}(\mathbf{x})}, \hspace{0.5cm} \overline{Z}_{R^c}(\overline{I}_{\partial R}, L_h) = \sum_{\substack{\overline{L} \in R^c \\ \left.\overline{L}\right|_{\partial R} = \overline{I}_{\partial R}}} c^{|\overline{L}|} t^{2\sum_{\mathbf{x}\in R^c}\varphi_{\overline{L}}(\mathbf{x})},
\end{equation}
where now $|\overline{L}|$ only includes the set of closed loops and not open strings, which correspond to loops in $I_{\partial R}$ that have been ``cut'' by the entanglement cut.  Due to the orthogonality relation
\begin{equation}
    \braket{Q(\psi_{R^c}(I'_{\partial R}, L'_h))}{Q(\psi_{R^c}(I_{\partial R}, L_h))} = \delta_{I, I'} \delta_{L_h, L_h'},
\end{equation}
the entanglement entropy is
\begin{equation}
    S(Q(\rho_R)) = -\sum_{I_{\partial R}, L_h} t^{2 \sum_{\mathbf{x}\in \partial R}|\varphi_{I, L_h}(\mathbf{x})|} c^{L(I)} \frac{\overline{Z}_{R^c}(I_{\partial R}, L_h) \overline{Z}_R(I_{\partial R}, L_h)}{\overline{Z}} \log \left(\sum_{L_h} t^{2 \sum_{\mathbf{x}\in \partial R}|\varphi_{I, L_h}(\mathbf{x})|} c^{L(I)} \frac{\overline{Z}_{R^c}(I_{\partial R}, L_h) \overline{Z}_R(I_{\partial R}, L_h)}{\overline{Z}}\right)
\end{equation}
Using the fact that $\overline{Z} = Z$ and $\overline{Z}_R(\overline{I}_{\partial R}, L_h) = Z_R(\overline{I}_{\partial R}, L_h)$, we arrive at the desired result $S(Q(\rho_R)) = -\sum_{\overline{I}_{\partial R}} p_R(\overline{I}_{\partial R}) \log p_R(\overline{I}_{\partial R})$.

In passing, we note that the state $\ket{Q(\psi)}$ is also the ground state of a local frustration-free parent Hamiltonian (which despite containing only a single color of loops carries a dependence on the parameter $c$ due to the way that different loop configurations are weighted). 
Let us briefly describe the parent Hamiltonian for this wavefunction.
Because there is only a single color of loops which appears in $\ket{Q(\psi)}$, the color label in the processes listed in Fig.~\ref{fig:processes} is no longer needed.  However, the first process in Fig.~\ref{fig:processes} changes the number of loops by $1$, while the second and third processes do not change the number of loops; therefore, we must add some dependence on $|c|$ to the first process.  This gives a parent Hamiltonian $H(t,c) = H_{\text{con}} + H_{\text{kin}}(t,c)$, with
\begin{align}
    H_{\text{kin}}(t,c) = &\sum_{v} \mathcal{A}_v(M_{1}^{(0)}) \mathcal{A}_v(M_{1}^{(1)})\mathcal{P}_v\left(t \sqrt{c} \ket{M_{1}^{(0)}} - \ket{M_{1}^{(1)}}\right) \nonumber\\ &+ \sum_{v,i}\mathcal{A}_v(M_{2,i}^{(0)}) \mathcal{A}_v(M_{2,i}^{(1)})\mathcal{P}_v\left(t \ket{M_{2,i}^{(0)}} - \ket{M_{2,i}^{(1)}}\right) \nonumber\\ &+ \sum_{v,i}\mathcal{A}_v(M_{3,i}^{(0)}) \mathcal{A}_v(M_{3,i}^{(1)})\mathcal{P}_v\left(t \ket{M_{3,i}^{(0)}} - \ket{M_{3,i}^{(1)}}\right).
\end{align}

\medskip 

We return to the calculation of the TEE.  Having reduced the computation of the entanglement entropy to that of $\langle \ell \rangle$, we next introduce some inclusion-exclusion-type identities that will be useful to compute the TEE. First, take the Kitaev-Preskill partition for the topological entanglement entropy.  For this choice of partition, we claim that
\begin{equation}
    \text{TEE}_{\text{KP}}(\rho) - \text{TEE}_{\text{KP}}(Q(\rho)) = (\langle \ell \rangle_{A} + \langle \ell \rangle_{B} + \langle \ell \rangle_{C} - \langle \ell \rangle_{AB} - \langle \ell \rangle_{BC} - \langle \ell \rangle_{AC} + \langle \ell \rangle_{ABC}) \log c.
\end{equation}
First note that any loop contained entirely within $ABC$ will not contribute to the TEE.  This can be verified on a case-by-case basis.  For example, if a loop is entirely in $AB$ but not entirely in $A$ or $B$, then this loop contributes to $\langle \ell \rangle_A$, $\langle \ell \rangle_{B}$, $\langle \ell \rangle_{AC}$, and $\langle \ell \rangle_{BC}$, and therefore cancels out.  All other possibilities can be checked to work out similarly. 

Therefore, the only loops that can contribute nontrivially are loops that intersect the boundary of $ABC$.  We now define $\langle \mathcal{L} \rangle_R$ for $R \in \{A,B,C\}$ to be the number of loops that pass through both $R$ and $(ABC)^c$, but do not pass through $ABC \setminus R$.  We define $\langle \mathcal{L} \rangle_{R_1, R_2, \cdots}$ for $R_1 \in \{A,B,C\}, R_2 \in \{A,B,C\}, \cdots$ to be the number of loops that passes into each of $R_1, R_2, \cdots$ and $(ABC)^c$, but do not pass through $ABC \setminus (R_1\cup R_2 \cup \cdots)$.  For succinctness, we will refer to $\langle \mathcal{L} \rangle_{R_1, R_2, \cdots}$ as $\langle \mathcal{L} \rangle_{R_1R_2\cdots}$.  Then, we have the identities
\begin{align}
    \langle \ell \rangle_A &= \langle \mathcal{L} \rangle_A + \langle \mathcal{L} \rangle_{AC} + \langle \mathcal{L} \rangle_{AB} + \langle \mathcal{L} \rangle_{ABC} \nonumber \\
    \langle \ell \rangle_B &= \langle \mathcal{L} \rangle_B + \langle \mathcal{L} \rangle_{BC} + \langle \mathcal{L} \rangle_{AB} + \langle \mathcal{L} \rangle_{ABC} \nonumber \\
    \langle \ell \rangle_C &= \langle \mathcal{L} \rangle_C + \langle \mathcal{L} \rangle_{AC} + \langle \mathcal{L} \rangle_{BC} + \langle \mathcal{L} \rangle_{ABC}
\end{align}
for single regions, 
\begin{align}
    \langle \ell \rangle_{AB} &= \langle \mathcal{L} \rangle_A + \langle \mathcal{L} \rangle_B + \langle \mathcal{L} \rangle_{AB} + \langle \mathcal{L} \rangle_{BC} + \langle \mathcal{L} \rangle_{AC} + \langle \mathcal{L} \rangle_{ABC} \nonumber \\
    \langle \ell \rangle_{BC} &= \langle \mathcal{L} \rangle_B + \langle \mathcal{L} \rangle_C + \langle \mathcal{L} \rangle_{AB} + \langle \mathcal{L} \rangle_{BC} + \langle \mathcal{L} \rangle_{AC} + \langle \mathcal{L} \rangle_{ABC} \nonumber \\
    \langle \ell \rangle_{AC} &= \langle \mathcal{L} \rangle_A + \langle \mathcal{L} \rangle_C + \langle \mathcal{L} \rangle_{AB} + \langle \mathcal{L} \rangle_{BC} + \langle \mathcal{L} \rangle_{AC} + \langle \mathcal{L} \rangle_{ABC}
\end{align}
for pairs of regions, and for all three regions
\begin{equation}
    \langle \ell \rangle_{ABC} = \sum_{S \subseteq \{A,B,C\}} \langle \mathcal{L} \rangle_{S}.
\end{equation}
Using these identities we can now show that
\begin{equation}
    \text{TEE}_{\text{KP}}(\rho) - \text{TEE}_{\text{KP}}(Q(\rho)) = \langle \mathcal{L} \rangle_{ABC} \log c.
\end{equation}
This completes the proof of Theorem 1. 

To prove Theorem 2, we compute the topological entanglement entropy using the Levin-Wen partition scheme using an analogous logic as above.  We find that this definition of the TEE gives
\begin{equation}
    \text{TEE}_{\text{LW}}(\rho) - \text{TEE}_{\text{LW}}(Q(\rho)) =  \langle \mathcal{L} \rangle_{AC} \log c. 
\end{equation}
A visual depiction of the loops $\langle \mathcal{L} \rangle_{ABC}$ and $\langle \mathcal{L} \rangle_{AC}$ are shown in Figure~\ref{fig:TEE}.

The utility of these theorems is that they allow us to compute the TEE using simple geometric arguments and an analysis of the value of $\langle \mathcal{L}\rangle_{ABC}$ for the typical configurations that contribute to the TEE. Consider for example the KP prescription. If the expected number of loops crossing $ABC$ and its complement is proportional to the size of the partition, then one of $\rho$ and $Q(\rho)$ must have anomalous topological entanglement entropy. 
That this indeed occurs when $t>1$ was argued for non-rigorously in the main text below the presentation of Theorem 2.
% This is visualized in Figure~\ref{fig:TEE}.

\section{A more rigorous guarantee of anomalous TEE}\label{app:rigorous_tee}

In this section, we provide a more rigorous guarantee for anomalous TEE (focusing on the KP prescription) which goes beyond an analysis of most likely configurations.  First, we recapitulate what we mean by most probable configurations in the ground state.  Namely, this refers to loop configurations with the largest amplitude $\braket{L}{\psi}$.  On the $L\times L$ square lattice, the maximum amplitude configuration is comprised of $L/2$ concentric square loops forming a single ``mountain.''  Call the set of such loops $L$.  Each loop $\ell \in L$ can come in one of two colors, and the color choice does not affect the amplitude. If we define $\ket{\psi_{\rm mp}}$ as a uniform superposition over only these most probable states, then
\begin{equation}
    \ket{\psi_{\text{mp}}} \propto \bigotimes_{\ell \in L} (\ket{\textcolor{red}{\ell_r}} + \ket{\textcolor{blue}{\ell_b}})
\end{equation}
which is a tensor product of $L/2$ ``cat states.''  The non-local area law is immediate from this state, as the bipartite entanglement of $\ket{\psi_{\text{mp}}}$ has area law entanglement, but this entanglement is due to non-local correlations originating from the cat states. That the entanglement of $\ket{\psi_{\rm mp}}$ is close to the entanglement of the true ground state $\ket{\psi}$ follows from the fact that typical configurations are highly concentrated about this most probable configuration, as the amplitude of a configuration of loops with maximum height $(1-\epsilon)L/2$ (at the center of the lattice) is suppressed by a factor of $\exp(-\text{poly}(\epsilon) L^3)$ (implying that the trace distances of $\ket{\psi}$ and $\ket{\psi_{\rm mp}}$ are exponentially close).

We may quantify this non-local area law by computing the topological entanglement entropy using the Kitaev-Preskill prescription.  Concretely, suppose the region $ABC$ is elliptical with ellipticity $\epsilon$, and has a center coinciding with the center of the $L\times L$ square lattice.  The typical  ``single mountain'' configuration that dominates the weight of the ground state will have the property that $\mathcal{L}_{ABC} = \epsilon' \ell$ for some $\epsilon' = \text{poly}(\epsilon)$.  More specifically, suppose the lengths of the $AB$ and $BC$ partitions is $\ell(1+ \epsilon'/2)$ and the length of the $AC$ partition is $\ell(1- \epsilon'/2)$; then $\mathcal{L}_{ABC} = \epsilon' \ell$.  We denote the volume of the maximum mountain configuration by $V_{\text{max}}$, which is $O(L^3)$.

Consider a configuration of loops where $\mathcal{L}_{ABC} \leq r \epsilon' \ell$ for some $r < 1$.  We now ask what the maximum volume one can attain given this constraint on configurations.  

We conjecture that to maximize the volume (or achieve a value close to the maximum volume) given this constraint, we mimic the typical configuration featuring a mountain of concentric loops up to radius $\ell(1+ \epsilon'(1-2r)/2)$.  Below this radius, one must choose loops to maximize the volume while creating no loops crossing into all four regions.  The total volume of such a configuration is $V_{\text{max}} - \delta \ell^3$ for some $\delta$.  Therefore,
\begin{align}
    \langle \mathcal{L} \rangle_{ABC} &\geq r \epsilon' \ell\left(1 -\mathbb{P}(\mathcal{L} \leq r \epsilon' \ell)\right)\nonumber \\
    &\geq r \epsilon' \ell\left(1 -\frac{\sum_{C: \mathcal{L} \leq r \epsilon' \ell} t^{V(C)}}{\sum_{C} t^{V(C)}}\right)\nonumber \\
    &\geq r \epsilon' \ell\left(1 -\frac{\sum_{C: \mathcal{L} \leq r \epsilon' \ell} t^{V_{\text{max}}-\delta \ell^3}}{t^{V_{\text{max}}}}\right)\nonumber \\
    &\geq r \epsilon' \ell\left(1 - (\text{dim}\,\mathcal{H})^{2L^2} t^{-\delta \ell^3}\right).
\end{align}
Therefore, $\langle \mathcal{L} \rangle_{ABC} = O(\epsilon' \ell)$ for $\ell = \Omega(L^{2/3})$.  In particular, it is possible to have anomalous TEE for sizes $\ell$ where $\ell/L \to 0$ in the thermodynamic limit.  However, given that these bounds are very loose, it is possible that the anomalous TEE can occur over much smaller length scales than the bound provides.

\section{Boundary conditions for the biased colored loop model}\label{app:bcl_bconds}

Thus far, we have been considering the biased colored loop model on open boundary conditions, where the height fields are zero at the boundaries (this constraint is enforced by energetically penalizing any arrow configurations on the links at the border of the lattice).  However, crucially, the model does not need to be defined on open boundary conditions.  Consider defining the model on a sphere.  Before proceeding we must discuss how to define this model on a triangulatable manifold, as we have been restricted to the square lattice in our analysis thus far.  For any triangulation, we may define the standard dual lattice formed by sites on the centers of triangles; the dual lattice is trivalent by construction and height fields live on its sites.  First we define a local Hilbert space $\mathcal{H} = \text{span}\{\ket{\text{---}}, \ket{\upframe_c}, \ket{\downframe_c}\}$, and similarly define $H_{\text{con}}$ so that it projects onto no colored bonds or two identically colored bonds on each site.  In a similar way to the construction on a square lattice, we may define the moves $M_{0,c}, M_{1,c,i}, M_{2,c,i}$
\begin{equation}
\includegraphics[scale=0.25]{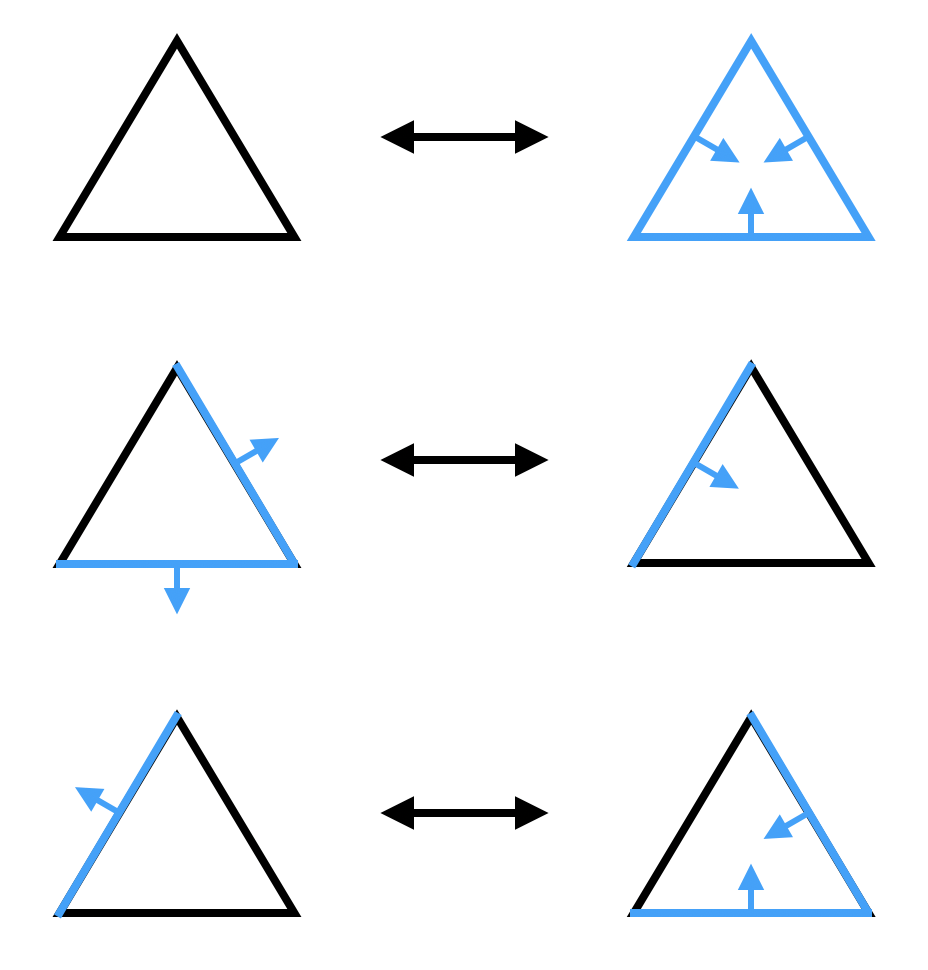}
\end{equation}
where the relative ratio of the right configuration and left configuration amplitudes is $t$.  This is because going from the left to the right configuration, the total volume (i.e. the sum $\sum_{\mathbf{x}} \varphi(\mathbf{x})$ where $\mathbf{x}$ is defined on the dual lattice) increases by 1.  In a similar manner to the biased colored loop model on the square lattice, we can define a frustration free Hamiltonian $H_{\text{kin}}$ such that the superposition of volume weighted loop configurations on the triangulation forms the ground state of $H_{\text{kin}} + H_{\text{con}}$.  

On a sphere of area $L^2$, which can be triangulated, the ground state is then still a superposition of framed loop configurations, but the ground state will not necessarily be unique. Let us consider the ground state connected to the empty sector with no loops.  The typical configuration in the ground state will still be a single mountain of loops when $t > 1$, but translates of this mountain are equivalent typical configurations, and so the number of such configurations is now $O(L^2)$ (of course, this extensive degeneracy will generically be lifted by any translation-breaking perturbation).

We ask whether on the sphere there is still an anomalous TEE.  To proceed, we consider the typical state
\begin{equation}
    \ket{\Omega} \approx \frac{1}{L} \sum_{(x,y)} \ket{M_{(x,y)}}
\end{equation}
where $\ket{M_{(x,y)}}$ is a single mountain of loops centered around coordinates $(x,y)$.  Note that the loops are colored, so $\ket{M_{(x,y)}}$ is actually an equal weight superposition of all possible color assignments to the mountain configuration (i.e. a tensor product of cat states as discussed in the previous Appendix).  We now apply the loop counting formula to this state and compute its TEE.

We will consider a Kitaev-Preskill partition that is circular with linear size $\ell$. Suppose that a given mountain configuration contributing to the sum in $|\Omega\rangle$ has its center shifted by some amount $\delta \ell$ away from the center of the Kitaev-Preskill partition; then anomalous TEE remains for this particular mountain configuration.  Therefore, for a fixed location of the Kitaev-Preskill partition, there are $O(\ell^2)$ mountains which possess anomalous TEE.  Since the total number of mountains is $L^2$, the TEE is roughly
\begin{equation}
    \text{TEE} \sim \frac{\epsilon \ell^2}{L^2} \cdot O(\ell) = O\left(\frac{\ell^3}{L^2}\right).
\end{equation}
This is $O(\ell)$ only when $\ell = O(L)$ and therefore anomalous TEE is expected to occur when the size of the region is extensively large. Thus the global topology of the manifold on which our model is defined plays an important role in determining the sizes of regions which can manifest an appreciably large anomalous TEE.

Finally, we also discuss the case when the Hamiltonian is placed on a torus or similar manifold with nontrivial homology.  In this case, two configurations that differ by some number of loops wrapping around homologically nontrivial cycles cannot be connected to one another via the terms in the Hamiltonian.  Therefore, ground states can be labelled by a canonical set of homologically nontrivial framed loops, resulting in an exponentially large ground state degeneracy.  

A given ground state can be viewed as a superposition of configurations with some fixed number of homologically nontrivial loops.  These loops are moveable, but cannot merge with one another -- therefore, regions between adjacent loops are disconnected from one another and can host their own framed loop configurations. 

\section{Subsystem symmetry breaking versus biased colored loop model}\label{app:subsystem}

In this Appendix, we inquire between whether there can be a simpler model whose ground state can exhibit anomalous topological entanglement entropy.  We now provide an argument that this phenomenon is expected to occur for systems with rigid subsystem symmetries, but that the biased colored loop model is more robust.

As an example, we consider the Xu-Moore model~\cite{PhysRevLett.93.047003}, defined on a square lattice with spins on the sites of the lattice:
\begin{equation}
    H = -J \sum_{\square} Z_1 Z_2 Z_3 Z_4 + h \sum_i X_i.
\end{equation}
This model has a rigid subsystem symmetry, with symmetry action given by
\begin{equation}
Q_{r_k/c_k} = \bigotimes_{i \in {r_k/c_k}} X_{i} 
\end{equation}
where $r_k$ and $c_k$ label rows and/or columns of the square lattice ($k = 1,2,\cdots,L$).  This is a symmetry operator by virtue of the fact that $[H, Q_{r_k}] = [H, Q_{r_c}] = 0$.  When $h = 0$, the ground state subspace is degenerate.  We can label the degenerate ground states by two vectors $\vec{R}$ and $\vec{C}$, both of which are elements of $\{0,1\}^L$.  We can define 
\begin{equation}
    Q_{\vec{R}} = \prod_{k=1}^L Q_{r_k}^{R_k}, \hspace{0.7cm} Q_{\vec{C}} = \prod_{k=1}^L Q_{c_k}^{C_k}
\end{equation}
which is the product of symmetry operators defined by $\vec{R}$ or $\vec{C}$.  We can label a ground state by
\begin{equation}
    \ket{\Omega_{\vec{R},\vec{C}}} = Q_{\vec{R}} Q_{\vec{C}} \bigotimes_i \ket{\downarrow_i}.
\end{equation}
We also have the following constraint on the symmetry operators
\begin{equation}
    \prod_{k=1}^L Q_{r_k} Q_{c_k} = 1,
\end{equation}
which imposes a constraint on the possible values of $\vec{R}$ and $\vec{C}$.  Under a weak transverse field $h \ll J$, the true ground state becomes
\begin{equation}
    \ket{\Omega} \propto \sum_{\vec{R},\vec{C} \in \{0,1\}^L}\ket{\Omega_{\vec{R},\vec{C}}}.
\end{equation}
where the sum is over independent pairs of $\vec{R}$ and $\vec{C}$.  The ground state is a superposition of $2^{2L-1}$ states.  Any reduced density
matrix formed by tracing out half the system will necessarily have entropy upper bounded by $O(L)$, indicating area
law scaling.  However, this area law scaling is coming from non-local entanglement.  To see this, note that the hamming distance $\text{dist}(\Omega_{\vec{R}, \vec{C}},\Omega_{\vec{R}', \vec{C}'}) \geq O(L)$, meaning that the different classical configurations in the superposition are `far apart' and intuitively this state cannot possess local correlations.  Another way to see this is that along each row $r$, the classical configurations where $Q_r$ is applied and $Q_r$ is not applied have the same amplitude in the ground state.  These two configurations can be grouped together forming a cat state along $r$.  When a vertical cut is made, the entanglement entropy from this cat state is $O(1)$ but comes from non-local entanglement.  Therefore, we expect $\ket{\Omega}$ to have similar entanglement properties to a stack of 1D cat states.

These considerations imply that $\ket{\Omega}$ should also possess anomalous TEE.  However, the biased loop model appears to have advantages over this far simpler construction, apart from being a qualitatively different way of obtaining anomalous TEE:
\begin{enumerate}
    \item The biased colored loop model is more robust:  the example of ground states with subsystem symmetries is very fragile, since merely adding a boundary longitudinal field $\sum_i Z_i$ can lift the ground state degeneracy and destroy the entanglement.  This also is true if a longitudinal field is applied anywhere in the bulk.  This is however not true for the biased colored loop model.  When a longitudinal field is applied in some region $R$, it enforces a boundary condition on $\partial R$ that loops need to satisfy (for example, there can be no loops in $R$ if a chemical potential for empty bonds are added in $R$).  Therefore, loop configurations are unconstrained in $R^c$ and must satisfy boundary conditions on $\partial R$.  In $R^c$, the entanglement properties are essentially unchanged since the biased colored loop ground state can form in $R^c$.  Therefore, by applying a boundary field or a bulk field in some region $R$, the anomalous TEE in $R^c$ cannot be affected, unless there is significant fine-tuning in the choice of $R$.  The only way to destroy the anomalous TEE is to apply a longitudinal field in the entire system.
    \item Because the Hamming distance is large between classical configurations in the ground state, the ground state for systems with subsystem symmetry cannot be prepared through a local dynamical process; instead, measurements are likely needed to induce long range correlations.  The ground state of the biased colored loop model can be constructed from a local dynamical process, which provides a potential advantage in terms of state preparation (and also provides an interesting avenue to study local dynamics yielding phases which develop non-local entanglement).
\end{enumerate}

\section{Hamiltonian for the decorated biased loop model}\label{app:ham_decorated_bcl}

We now discuss a modified version of the biased colored loop model, which we call the \emph{decorated biased loop model}. This model is interesting because its ground state possesses a bipartite entanglement entropy $S(A)$ which scales as a nonzero power of $|A|$, with a volume law being obtained in a certain region of parameter space. 

In the analysis of this model we will restrict ourselves to loops of only a \emph{single color}; our analysis can however be easily generalized to the multi-color case. 
While the construction in the main text is defined on a direct square lattice, the following construction will be presented with respect to the \emph{dual} lattice.  To proceed, we introduce a Hilbert space
\begin{equation}
\mathcal{H} = \mathcal{H}_{\text{vac}} \oplus (\mathcal{H}_{\text{arr}} \otimes \mathcal{H}_{\text{mot}})
\end{equation}
where for the horizontal links
\begin{equation}
\mathcal{H}_{\text{vac}} = \text{span}\{\ket{-}\}, \hspace{0.5cm} \mathcal{H}_{\text{arr}} = \text{span}\{\ket{\leftarrow}, \ket{\rightarrow}\}
\end{equation}
and for the vertical links
\begin{equation}
\mathcal{H}_{\text{vac}} = \text{span}\{\ket{|}\}, \hspace{0.5cm} \mathcal{H}_{\text{arr}} = \text{span}\{\ket{\uparrow}, \ket{\downarrow}\}
\end{equation}
and the remaining Hilbert space is given by
\begin{equation}
\mathcal{H}_{\text{mot}} = \text{span}\{\ket{0}, \ket{(_d}, \ket{)_d}\}.
\end{equation}
In the main text, we used the notation $\text{span}\{\ket{0}, \ket{\langle_d}, \ket{\rangle_d}\}$ but here we will revert to using round parentheses.  In this notation, $d$ is a label for color of the parentheses that decorate the arrow variables, with the $\mathcal{H}_{\rm mot}$ tensor factor amounting to the introduction of a $2|d|+1$-dimensional internal degree of freedom at each link of the dual lattice. For now we assume $|d| \geq 2$; the $|d| = 1$ case will be discussed briefly in a separate section.  Note that we may set $\mathcal{H}_h \cong \mathcal{H}_v$ by identifying right and up arrows and left and down arrows, but for clarity of presentation we will keep the above notation. 

As before, we constrain the Hilbert space to the following plaquette configurations: 
\begin{equation}
\begin{array}{ccccc}
\begin{tikzpicture}[baseline={([yshift=-.5ex]current bounding box.center)},vertex/.style={anchor=base,circle,fill=black!25,minimum size=18pt,inner sep=2pt}]
    \draw (0,0) -- (0.8,0) -- (0.8,0.8) -- (0,0.8) -- (0,0); 
  \end{tikzpicture} & \begin{tikzpicture}[baseline={([yshift=-1.5ex]current bounding box.center)},vertex/.style={anchor=base,circle,fill=black!25,minimum size=18pt,inner sep=2pt}]
    \draw (0,0) -- (0.8,0) -- (0.8,0.8) -- (0,0.8) -- (0,0); 
    \draw[->] (0.8,0.8)--(0.4,0.8) node[above] {$\ast$};
    \draw[->] (0,0)--(0,0.4) node[left] {$\ast$};
    \node[right] at (0.8,0.4) {$\text{ }$};
  \end{tikzpicture} & \begin{tikzpicture}[baseline={([yshift=-1.5ex]current bounding box.center)},vertex/.style={anchor=base,circle,fill=black!25,minimum size=18pt,inner sep=2pt}]
    \draw (0,0) -- (0.8,0) -- (0.8,0.8) -- (0,0.8) -- (0,0); 
    \draw[->] (0.8,0)--(0.8,0.4) node[right] {$\ast$};
    \draw[->] (0,0.8)--(0.4,0.8) node[above] {$\ast$};
  \end{tikzpicture} & \begin{tikzpicture}[baseline={([yshift=1ex]current bounding box.center)},vertex/.style={anchor=base,circle,fill=black!25,minimum size=18pt,inner sep=2pt}]
    \draw (0,0) -- (0.8,0) -- (0.8,0.8) -- (0,0.8) -- (0,0); 
    \draw[->] (0,0)--(0.4,0) node[below] {$\ast$};
    \draw[->] (0.8,0.8)--(0.8,0.4) node[right] {$\ast$};
  \end{tikzpicture} & \begin{tikzpicture}[baseline={([xshift=1ex, yshift=1ex]current bounding box.center)},vertex/.style={anchor=base,circle,fill=black!25,minimum size=18pt,inner sep=2pt}]
    \draw (0,0) -- (0.8,0) -- (0.8,0.8) -- (0,0.8) -- (0,0); 
    \draw[->] (0.8,0)--(0.4,0) node[below] {$\ast$};
    \draw[->] (0,0.8)--(0,0.4) node[left] {$\ast$};
  \end{tikzpicture}
  \end{array}
\end{equation}
as well as
\begin{equation}
\begin{array}{cccc}
  \begin{tikzpicture}[baseline={([yshift=1.5ex]current bounding box.center)},vertex/.style={anchor=base,circle,fill=black!25,minimum size=18pt,inner sep=2pt}]
    \draw (0,0) -- (0.8,0) -- (0.8,0.8) -- (0,0.8) -- (0,0); 
    \draw[->] (0,0)--(0.4,0) node[below] {$\ast$};
    \draw[->] (0,0.8)--(0.4,0.8) node[above] {$\ast$};
  \end{tikzpicture} & \begin{tikzpicture}[baseline={([yshift=1.5ex]current bounding box.center)},vertex/.style={anchor=base,circle,fill=black!25,minimum size=18pt,inner sep=2pt}]
    \draw (0,0) -- (0.8,0) -- (0.8,0.8) -- (0,0.8) -- (0,0); 
    \draw[->] (0.8,0)--(0.8,0.4) node[right] {$\ast$};
    \draw[->] (0,0)--(0,0.4) node[left] {$\ast$};
  \end{tikzpicture} & \begin{tikzpicture}[baseline={([yshift=2ex]current bounding box.center)},vertex/.style={anchor=base,circle,fill=black!25,minimum size=18pt,inner sep=2pt}]
    \draw (0,0) -- (0.8,0) -- (0.8,0.8) -- (0,0.8) -- (0,0); 
    \draw[->] (0,0)--(0.4,0) node[below] {$\ast$};
    \draw[->] (0.8,0)--(0.8,0.4) node[right] {$\ast$};
    \draw[->] (0,0.8)--(0.4,0.8) node[above] {$\ast$};
    \draw[->] (0,0)--(0,0.4) node[left] {$\ast$};
  \end{tikzpicture} & \begin{tikzpicture}[baseline={([yshift=2ex]current bounding box.center)},vertex/.style={anchor=base,circle,fill=black!25,minimum size=18pt,inner sep=2pt}]
    \draw (0,0) -- (0.8,0) -- (0.8,0.8) -- (0,0.8) -- (0,0); 
    \draw[->] (0.8,0)--(0.4,0) node[below] {$\ast$};
    \draw[->] (0.8,0)--(0.8,0.4) node[right] {$\ast$};
    \draw[->] (0.8,0.8)--(0.4,0.8) node[above] {$\ast$};
    \draw[->] (0,0)--(0,0.4) node[left] {$\ast$};
  \end{tikzpicture}\\
\end{array}
\end{equation}
as well as configurations where the arrows are reversed.  In this notation, $\ast$ is a stand-in for either no decoration, $(_d$, or $)_d$.  Based on these constraints, we can define Hamiltonian terms $\widetilde{H}_{\text{con}, p}$ and $\widetilde{V}_p$ in an analogous way to the terms $H_{\text{con}, p}$, $V_p$ that appear in the undecorated biased colored loop model.  The processes are identical to those in Equation~\ref{eqn:processes}, and we may analogously define a Hamiltonian
\begin{align}
    \widetilde{H}_{\text{loop}}(t) =  \widetilde{H}_{\text{con}} + \widetilde{H}_{\text{kin}}(t) = &\sum_{v} \widetilde{\mathcal{A}}_v(M_{1}^{(0)}) \widetilde{\mathcal{A}}_v(M_{1}^{(1)})\mathcal{P}_v\left(t \ket{M_{1}^{(0)}} - \ket{M_{1}^{(1)}}\right) \nonumber\\ &+ \sum_{v,i}\widetilde{\mathcal{A}}_v(M_{2,i}^{(0)}) \widetilde{\mathcal{A}}_v(M_{2,i}^{(1)})\mathcal{P}_v\left(t \ket{M_{2,i}^{(0)}} - \ket{M_{2,i}^{(1)}}\right) \nonumber\\ &+ \sum_{v,i}\widetilde{\mathcal{A}}_v(M_{3,i}^{(0)}) \widetilde{\mathcal{A}}_v(M_{3,i}^{(1)})\mathcal{P}_v\left(t \ket{M_{3,i}^{(0)}} - \ket{M_{3,i}^{(1)}}\right).
\end{align}
We must also discuss the constraint that we enforce on the $\ast$ degrees of freedom.  To this end, we define additional processes associated with the decorating Hilbert space:
\begin{align}
M_{4,1,d,i} &= \begin{tikzpicture}[baseline={([yshift=-1.5ex]current bounding box.center)},vertex/.style={anchor=base,circle,fill=black!25,minimum size=18pt,inner sep=2pt}]
    \draw (0,0) -- (0,0.8) -- (0.8,0.8); 
    \draw[->] (0.8,0.8)--(0.4,0.8) node[above] {$\text{ }$};
    \draw[->] (0,0)--(0,0.4) node[left] {$\text{ }$};
    \node[right] at (0.8,0.4) {$\text{ }$};
  \end{tikzpicture} \Longleftrightarrow \begin{tikzpicture}[baseline={([yshift=-1.5ex]current bounding box.center)},vertex/.style={anchor=base,circle,fill=black!25,minimum size=18pt,inner sep=2pt}]
    \draw (0,0) -- (0,0.8) -- (0.8,0.8);  
    \draw[->] (0.8,0.8)--(0.4,0.8) node[above] {$)_d$};
    \draw[->] (0,0)--(0,0.4) node[left] {$(_d$};
    \node[right] at (0.8,0.4) {$\text{ }$};
  \end{tikzpicture}\hspace{0.75cm}
M_{4,2,d,i} = \begin{tikzpicture}[baseline={([yshift=-1.5ex]current bounding box.center)},vertex/.style={anchor=base,circle,fill=black!25,minimum size=18pt,inner sep=2pt}]
    \draw (0,0) -- (0,0.8) -- (0.8,0.8); 
    \draw[->] (0.8,0.8)--(0.4,0.8) node[above] {$\phantom{(}$};
    \draw[->] (0,0)--(0,0.4) node[left] {$)_d$};
    \node[right] at (0.8,0.4) {$\text{ }$};
  \end{tikzpicture} \Longleftrightarrow \begin{tikzpicture}[baseline={([yshift=-1.5ex]current bounding box.center)},vertex/.style={anchor=base,circle,fill=black!25,minimum size=18pt,inner sep=2pt}]
    \draw (0,0) -- (0,0.8) -- (0.8,0.8);  
    \draw[->] (0.8,0.8)--(0.4,0.8) node[above] {$)_d$};
    \draw[->] (0,0)--(0,0.4) node[left] {$\phantom{(}$};
    \node[right] at (0.8,0.4) {$\text{ }$};
  \end{tikzpicture}\hspace{0.75cm}
M_{4,3,d,i} = \begin{tikzpicture}[baseline={([yshift=-1.5ex]current bounding box.center)},vertex/.style={anchor=base,circle,fill=black!25,minimum size=18pt,inner sep=2pt}]
    \draw (0,0) -- (0,0.8) -- (0.8,0.8); 
    \draw[->] (0.8,0.8)--(0.4,0.8) node[above] {$(_d$};
    \draw[->] (0,0)--(0,0.4) node[left] {$\phantom{(}$};
    \node[right] at (0.8,0.4) {$\text{ }$};
  \end{tikzpicture} \Longleftrightarrow \begin{tikzpicture}[baseline={([yshift=-1.5ex]current bounding box.center)},vertex/.style={anchor=base,circle,fill=black!25,minimum size=18pt,inner sep=2pt}]
    \draw (0,0) -- (0,0.8) -- (0.8,0.8);  
    \draw[->] (0.8,0.8)--(0.4,0.8) node[above] {$\phantom{(}$};
    \draw[->] (0,0)--(0,0.4) node[left] {$(_d$};
    \node[right] at (0.8,0.4) {$\text{ }$};
  \end{tikzpicture}\nonumber\\
M_{5,1,d,i} &= \begin{tikzpicture}[baseline={([yshift=-1.5ex]current bounding box.center)},vertex/.style={anchor=base,circle,fill=black!25,minimum size=18pt,inner sep=2pt}]
    \draw (0,0) -- (0,0.8) -- (0.8,0.8); 
    \draw[->] (0,0.8)--(0.4,0.8) node[above] {$\phantom{(}$};
    \draw[->] (0,0.8)--(0,0.4) node[left] {$\phantom{(}$};
    \node[right] at (0.8,0.4) {$\text{ }$};
  \end{tikzpicture} \Longleftrightarrow \begin{tikzpicture}[baseline={([yshift=-1.5ex]current bounding box.center)},vertex/.style={anchor=base,circle,fill=black!25,minimum size=18pt,inner sep=2pt}]
    \draw (0,0) -- (0,0.8) -- (0.8,0.8);  
    \draw[->] (0,0.8)--(0.4,0.8) node[above] {$(_d$};
    \draw[->] (0,0.8)--(0,0.4) node[left] {$)_d$};
    \node[right] at (0.8,0.4) {$\text{ }$};
  \end{tikzpicture}\hspace{0.75cm}
M_{5,2,d,i} = \begin{tikzpicture}[baseline={([yshift=-1.5ex]current bounding box.center)},vertex/.style={anchor=base,circle,fill=black!25,minimum size=18pt,inner sep=2pt}]
    \draw (0,0) -- (0,0.8) -- (0.8,0.8); 
    \draw[->] (0,0.8)--(0.4,0.8) node[above] {$)_d$};
    \draw[->] (0,0.8)--(0,0.4) node[left] {$\phantom{(}$};
    \node[right] at (0.8,0.4) {$\text{ }$};
  \end{tikzpicture} \Longleftrightarrow \begin{tikzpicture}[baseline={([yshift=-1.5ex]current bounding box.center)},vertex/.style={anchor=base,circle,fill=black!25,minimum size=18pt,inner sep=2pt}]
    \draw (0,0) -- (0,0.8) -- (0.8,0.8);  
    \draw[->] (0,0.8)--(0.4,0.8) node[above] {$\phantom{(}$};
    \draw[->] (0,0.8)--(0,0.4) node[left] {$)_d$};
    \node[right] at (0.8,0.4) {$\text{ }$};
  \end{tikzpicture}\hspace{0.75cm}
M_{5,3,d,i} = \begin{tikzpicture}[baseline={([yshift=-1.5ex]current bounding box.center)},vertex/.style={anchor=base,circle,fill=black!25,minimum size=18pt,inner sep=2pt}]
    \draw (0,0) -- (0,0.8) -- (0.8,0.8); 
    \draw[->] (0,0.8)--(0.4,0.8) node[above] {$\phantom{(}$};
    \draw[->] (0,0.8)--(0,0.4) node[left] {$(_d$};
    \node[right] at (0.8,0.4) {$\text{ }$};
  \end{tikzpicture} \Longleftrightarrow \begin{tikzpicture}[baseline={([yshift=-1.5ex]current bounding box.center)},vertex/.style={anchor=base,circle,fill=black!25,minimum size=18pt,inner sep=2pt}]
    \draw (0,0) -- (0,0.8) -- (0.8,0.8);  
    \draw[->] (0,0.8)--(0.4,0.8) node[above] {$(_d$};
    \draw[->] (0,0.8)--(0,0.4) node[left] {$\phantom{(}$};
    \node[right] at (0.8,0.4) {$\text{ }$};
  \end{tikzpicture}\nonumber\\
  M_{6,1,d,i} &= \begin{tikzpicture}[baseline={([yshift=-1.5ex]current bounding box.center)},vertex/.style={anchor=base,circle,fill=black!25,minimum size=18pt,inner sep=2pt}]
    \draw (0,0) -- (0.8,0);
    \draw (0.8,0.8) -- (0,0.8); 
    \draw[->] (0.8,0.8)--(0.4,0.8) node[above] {$\text{ }$};
    \draw[->] (0.8,0)--(0.4,0) node[below] {$\text{ }$};
    \node[left] at (0,0.4) {$\text{ }$};
    \node[right] at (0.8,0.4) {$\text{ }$};
  \end{tikzpicture} \Longleftrightarrow \begin{tikzpicture}[baseline={([yshift=-1.5ex]current bounding box.center)},vertex/.style={anchor=base,circle,fill=black!25,minimum size=18pt,inner sep=2pt}]
    \draw (0,0) -- (0.8,0);
    \draw (0.8,0.8) -- (0,0.8); 
    \draw[->] (0.8,0.8)--(0.4,0.8) node[above] {$(_d$};
    \draw[->] (0.8,0)--(0.4,0) node[below] {$)_d$};
    \node[left] at (0,0.4) {$\text{ }$};
    \node[right] at (0.8,0.4) {$\text{ }$};
  \end{tikzpicture} \hspace{0.75cm}
  M_{6,2,d,i} = \begin{tikzpicture}[baseline={([yshift=-1.5ex]current bounding box.center)},vertex/.style={anchor=base,circle,fill=black!25,minimum size=18pt,inner sep=2pt}]
    \draw (0,0) -- (0.8,0);
    \draw (0.8,0.8) -- (0,0.8); 
    \draw[->] (0.8,0.8)--(0.4,0.8) node[above] {$\phantom{(}$};
    \draw[->] (0.8,0)--(0.4,0) node[below] {$(_d$};
    \node[left] at (0,0.4) {$\text{ }$};
    \node[right] at (0.8,0.4) {$\text{ }$};
  \end{tikzpicture} \Longleftrightarrow \begin{tikzpicture}[baseline={([yshift=-1.5ex]current bounding box.center)},vertex/.style={anchor=base,circle,fill=black!25,minimum size=18pt,inner sep=2pt}]
    \draw (0,0) -- (0.8,0);
    \draw (0.8,0.8) -- (0,0.8); 
    \draw[->] (0.8,0.8)--(0.4,0.8) node[above] {$(_d$};
    \draw[->] (0.8,0)--(0.4,0) node[below] {$\phantom{(}$};
    \node[left] at (0,0.4) {$\text{ }$};
    \node[right] at (0.8,0.4) {$\text{ }$};
  \end{tikzpicture} \hspace{0.75cm}
  M_{6,3,d,i} = \begin{tikzpicture}[baseline={([yshift=-1.5ex]current bounding box.center)},vertex/.style={anchor=base,circle,fill=black!25,minimum size=18pt,inner sep=2pt}]
    \draw (0,0) -- (0.8,0);
    \draw (0.8,0.8) -- (0,0.8); 
    \draw[->] (0.8,0.8)--(0.4,0.8) node[above] {$)_d$};
    \draw[->] (0.8,0)--(0.4,0) node[below] {$\phantom{(}$};
    \node[left] at (0,0.4) {$\text{ }$};
    \node[right] at (0.8,0.4) {$\text{ }$};
  \end{tikzpicture} \Longleftrightarrow \begin{tikzpicture}[baseline={([yshift=-1.5ex]current bounding box.center)},vertex/.style={anchor=base,circle,fill=black!25,minimum size=18pt,inner sep=2pt}]
    \draw (0,0) -- (0.8,0);
    \draw (0.8,0.8) -- (0,0.8); 
    \draw[->] (0.8,0.8)--(0.4,0.8) node[above] {$\phantom{(}$};
    \draw[->] (0.8,0)--(0.4,0) node[below] {$)_d$};
    \node[left] at (0,0.4) {$\text{ }$};
    \node[right] at (0.8,0.4) {$\text{ }$};
  \end{tikzpicture}
\end{align}
where the subscript $d$ indicates the color of the decorating parentheses, and the subscript $i$ indicates that other processes are related by $\pi/2$ rotations of the arrow configurations.  In these processes, no label next to an arrowed link indicates $\ket{0}$.  Notice that in the loop picture on the direct lattice, each bond of the loop is now decorated with a parenthesis, the processes above enforce the Motzkin constraints on the parentheses decorating each loop.  The framing of the loop is necessary because parenthesis matching is uni-directional (i.e. an open parenthesis must come before its matching closed parenthesis), and the framing provides a preferred direction along which parenthesis matching can be done.

\begin{figure}
    \centering
    \includegraphics[scale=0.12]{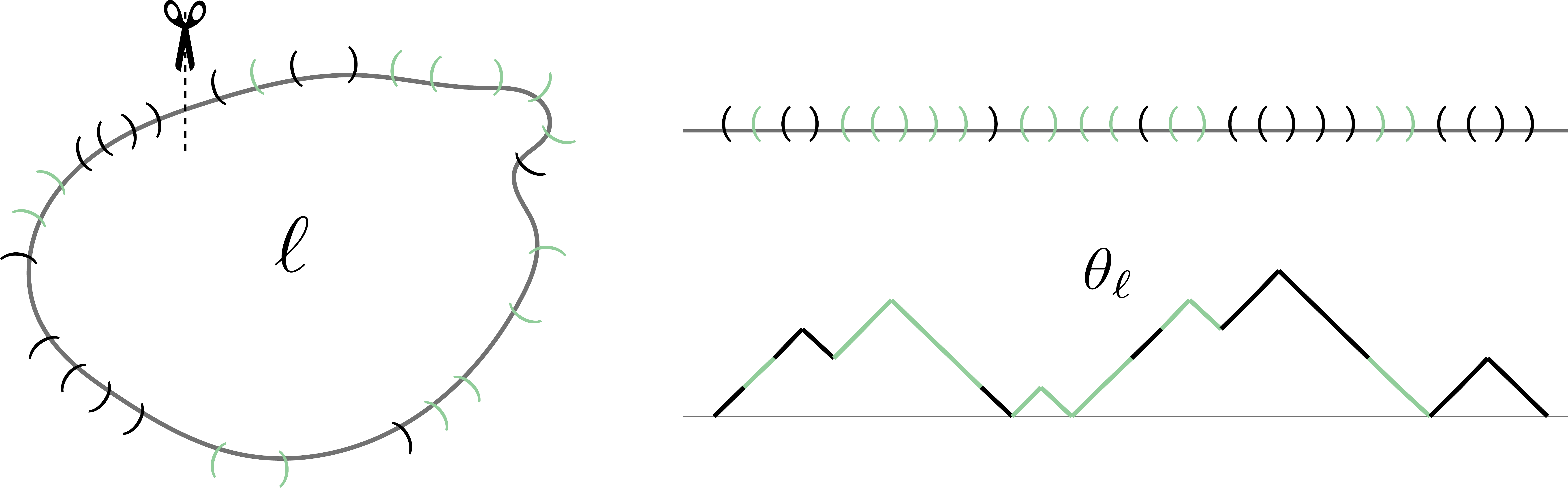}
    \caption{An illustration of a mapping between matched parentheses around a loop and a height field.  An open parenthesis corresponds to moving up-right, a closed parenthesis corresponds to moving down-right, and colors must match when open and closed parentheses are matched.}
    \label{fig:loopheightfield}
\end{figure}

Analogously to the Motzkin chain, for each loop $\ell$, one may define a secondary ``height'' field variable $\theta_{\ell}$ on the links of $\ell$, which measures the ``unmatchedness'' of the internal parenthesis degrees of freedom on that loop (an example is shown in Fig.~\ref{fig:loopheightfield}). Since the parentheses in the current context come in $d$ different colors, $\theta_\ell$ does not simply keep track of a total height. Rather, it keeps track of an entire collection of ``colored parentheses'' encountered along the loop. Thus the value of $\theta_\ell$ on a given link of $\ell$ is in fact a string consisting of (at most $|\ell|/2$) characters drawn from the set $\{ (_d\}$ of size $d$.

Explicitly, the value $\theta_{\ell,i}$ of the ``height field'' at a link $i$ of $\ell$ can be obtained by via the following algorithm. Starting from a pre-defined origin, one proceeds along $\ell$ in the direction of link $i$. When a $(_d$ is encountered, the symbol $(_d$ is appended to $\theta_{\ell,i}$. When a $)_d$ is encountered {\it and} the last character of $\theta_{\ell,i}$ is equal to $(_d$, the $(_d$ on the end of $\theta_{\ell,i}$ is erased. In this way $\theta_{\ell,i}$ is constructed to store the entire history of unmatched colored parentheses defined along $\ell$.  The origin is chosen so that the minimum height of $\theta_{\ell,i}$ is exactly zero.

As in the area deformed Motzkin chain, we may assign a weight of $u^{A(C_{\ell})}$ to configurations on $\ell$, where $A(C_{\ell})$ is a sum of the values of the height field around $\ell$ (by `value of the height field' at a given link, we mean the {\it length} of the string $\theta_\ell$ at that link).  We then introduce a Hamiltonian $H_{\text{mot}}(u)$, given by
\begin{align}
    H_{\text{mot}}(u) = &\sum_{v,j,d,i} \mathcal{P}_v\left(u \ket{M_{4,j,d,i}^{(0)}} - \ket{M_{4,j,d,i}^{(1)}}\right) + \sum_{v,j,d,i} \mathcal{P}_v\left(u \ket{M_{5,j,d,i}^{(0)}} - \ket{M_{5,j,d,i}^{(1)}}\right) + \sum_{v,j,d,i} \mathcal{P}_v\left(u \ket{M_{6,j,d,i}^{(0)}} - \ket{M_{6,j,d,i}^{(1)}}\right) .
\end{align}
This is a frustration-free Hamiltonian with a unique ground state (assuming $u \neq 1$) given by a superposition over matched parenthesis sequences on a given loop, where the amplitude for a given sequence of parentheses is area weighted.  Finally, we want to ensure that coloring is respected while matching parentheses.  The following configurations therefore must incur an energetic penalty:
\begin{equation}
\begin{array}{ccc}
  \begin{tikzpicture}[baseline={([yshift=1.5ex]current bounding box.center)},vertex/.style={anchor=base,circle,fill=black!25,minimum size=18pt,inner sep=2pt}]
    \draw (0,0) -- (0,0.8) -- (0.8,0.8); 
    \draw[->] (0.8,0.8)--(0.4,0.8) node[above] {$)_{d'}$};
    \draw[->] (0,0)--(0,0.4) node[left] {$(_d$};
    \node[left] at (0.4,0) {$\phantom{(_d}$};
    \node[right] at (0.8,0.4) {$\text{ }$};
  \end{tikzpicture} & \begin{tikzpicture}[baseline={([yshift=1.5ex]current bounding box.center)},vertex/.style={anchor=base,circle,fill=black!25,minimum size=18pt,inner sep=2pt}]
    \draw (0,0) -- (0,0.8) -- (0.8,0.8); 
    \draw[->] (0,0.8)--(0.4,0.8) node[above] {$(_d$};
    \draw[->] (0,0.8)--(0,0.4) node[left] {$)_{d'}$};
    \node[left] at (0.4,0) {$\phantom{(_d}$};
    \node[right] at (0.8,0.4) {$\text{ }$};
  \end{tikzpicture} & \begin{tikzpicture}[baseline={([yshift=2ex]current bounding box.center)},vertex/.style={anchor=base,circle,fill=black!25,minimum size=18pt,inner sep=2pt}]
    \draw (0,0) -- (0.8,0);
    \draw (0.8,0.8) -- (0,0.8); 
    \node[left] at (0,0.4) {$\text{ }$};
    \node[right] at (0.8,0.4) {$\text{ }$};
    \draw[->] (0.8,0)--(0.4,0) node[below] {$)_{d'}$};
    \draw[->] (0.8,0.8)--(0.4,0.8) node[above] {$(_d$};
  \end{tikzpicture}\\
\end{array}
\end{equation}
where $d \neq d'$.  Calling $\mathscr{D}_p$ the set of configurations of the above form at a plaquette $p$, and constructing the projector
\begin{equation}
H_{\text{mot, con}} = \sum_p \sum_{D \in \mathscr{D}_p} \ketbra{D}{D},
\end{equation}
we may write down the entire Hamiltonian as
\begin{equation}
H(t,u) = \widetilde{H}_{\text{con}} + H_{\text{mot, con}} + \widetilde{H}_{\text{kin}}(t) + H_{\text{mot}}(u).
\end{equation}

By construction, this Hamiltonian is frustration-free, one may write down the ground state exactly.  In particular, given a configuration of decorated loops $C$, call $L(C)$ the set of loops in the configuration, and $C_{\ell}$ the configuration restricted to loop $\ell$.  Then, the ground state wavefunction is
\begin{equation}
\ket{\psi} \propto \sum_{C} t^{V(C)} u^{\sum_{\ell \in L(C)} A(C_{\ell})} \ket{C}
\end{equation}
where $C$ indicates a configuration of nonintersecting loops on the dual lattice which are decorated with a matched sequence of parentheses.  An alternative rewriting of this ground state is presented in the main text, as well as a visual representation of this state.

\section{Entanglement scaling in the decorated biased loop model}

In this section, we will analyze the entanglement scaling of the Hamiltonian for the decorated biased loop model.  The decorated model has two parameters $t$ and $u$; $t$ controls the amplitude of a particular loop configuration, while $u$ controls the amplitude of a decorating Motzkin configuration around a loop.  We provide heuristic arguments for the following claims:
\begin{itemize}
\item When $u > 1$, the bipartite entanglement scaling of the ground state follows a volume law scaling $S(A) \sim L^2$.  We provide multiple different arguments for this claim.
\item When $u = 1$ and $t > 1$, the bipartite entanglement scaling of the ground state has a scaling violating the area law by a polynomial factor, $S(A) \sim L^{3/2}$.  We provide an argument that $L^{3/2}$ is a lower bound, and speculate that it is tight.
\end{itemize}
We also discuss the general structure of the phase diagram for the entanglement entropy as a function of $u$ and $t$.

\subsection{Entanglement scaling when $u > 1$}\label{app:volume_law}

In this section, we argue that the ground state of the decorated model constructed in the previous section has volume-law entanglement when $u>1$. As before, we will work on an $L\times L$ square lattice with open boundary conditions. 

We start by writing the ground state as
\begin{equation}
\ket{\psi} = \frac{1}{\sqrt{Z}}\sum_{C} t^{A(C)} u^{\sum_{\ell \in L(C)} A(C_{\ell})} \ket{C},
\end{equation}
where $\ell \in L$ denotes individual loops $\ell$ in loop configuration $L$, and the equation above is identical to the ground state presented in the main text for the decorated model. 
Roughly speaking, the amplitude for a given configuration is given by 
\begin{align}
    \langle C| \psi \rangle &\propto t^{\textrm{total volume under loops}} u^{\sum_\ell \textrm{total area under decorated loop } \ell}\\
    &\sim 
    t^{\textrm{total volume under loops}} u^{\sum_\ell (\textrm{length}(\ell))^2},
\end{align}
where in the second line we assumed that $u>1$, which will leads to the area under the decorated loop scaling quadratically in the length of the loop for typical configurations.
We now seek to understand what typical (saddle-point) configurations dominate in the ground state.  

The configuration corresponding to a single mountain of loops is actually \emph{not} the dominant saddle point configuration. To see this, first note that the amplitude of such a configuration is $t^{O(L^3)} u^{O(L^3)}$, as it contains $O(L)$ loops of length $O(L)$, each of which is weighted by a factor of $u^{O(L^2)}$ from the ``internal'' Motzkin degrees of freedom, and a factor of $t^{O(L^2)}$ from the area it encloses. 

Instead, consider a particular configuration corresponding to a {\it single} loop of length $O(L^2)$; any loop of this length will be wound around the $L\times L$ lattice in a densely-packed way somewhat reminiscent of a labyrinth.  Since the volume of this loop configuration is only $O(L^2)$, the $t$-dependence of its amplitude is only  $t^{O(L^2)}$.  However, the Motzkin chain decorating this long loop contributes amplitude $u^{O((L^2)^2)}$, so loops of this form are the ones which provide the dominant contributions to the ground state.  Note that this is true for $u > 1$ regardless of the value of $t$ as long as $u,t$ are both order 1, which is the regime we focus on in this section.  Intuitively, the weighting of the decorating Motzkin walks by the parameter $u>1$ means that the entropy is dominated by configurations with large entropy contained in the ``internal'' Motzkin degrees of freedom, rather than by configurations where the loops themselves are the deciding factor. 

To argue for volume law entanglement scaling, we therefore analyze the toy problem of computing the entanglement entropy of a wavefunction containing a single loop of length $O(L^2)$, on which a $u$-deformed Motzkin chain has been placed.
For simplicity we will focus on a straight cut that bipartitions the system into left and right halves $A,B$.

The fact that we get volume-law entanglement in this state should then be intuitively clear. Indeed, since the area-deformed colored Motzkin chain possesses a volume-law entangled ground state in 1d, the entanglement contributed by the large loop will scale as the length of the loop, viz. as $L^2$, thereby yielding a volume law when embedded in the $L\times L$ lattice.

We now try to make this intuition more precise.  Let us consider the case of a single loop of length $O(L^2)$, with the entanglement cut dividing the loop into regions $A_1, B_1, A_2, B_2, \cdots, A_k, B_k$, where the alternating labels $A$ and $B$ indicate that the assignments of the segments to ``Alice'' (left side of the bipartition) or ``Bob'' (right side) alternate.  Around this loop, the ground state forms a Motzkin chain which is volume-deformed with deformation parameter $u$.  The most probable configuration corresponds to a ``mountain'' of parentheses that look like $((((\cdots))))$.  Due to the coloring, a pair of matched parentheses corresponds to a Bell state, i.e. $\ket{\textcolor{red}{()}} + \ket{\textcolor{blue}{()}}$ (plus additional terms if there are more colors).  Therefore the most probable configuration assuming a fixed location for the mountain peak is
\begin{equation}
    \ket{\psi} \propto \bigotimes_{i=1}^{L^2/2} (\ket{\textcolor{red}{(_i )_{L^2-i}}} + \ket{\textcolor{blue}{(_i )_{L^2-i}}}). 
\end{equation}
However, the set of all most probable configurations include those where the peak of the mountain is translated by any amount. To break this degeneracy we define $\ket{\psi_i}$ to be the mountain configuration where the peak of the mountain is at location $i$.  The most probable state is $\ket{\psi_{\text{mp}}} \propto \sum_i \ket{\psi_i}$.  The goal is now to compute the entanglement entropy corresponding to $\rho_A = \text{Tr}_B\left(\ketbra{\psi_{\text{mp}}}{\psi_{\text{mp}}}\right)$; i.e. $S_A = -\text{Tr}\left(\rho_A \log \rho_A\right)$.

We argue that this state admits a nice Schmidt decomposition.  In a given mountain configuration on a loop, there will be a single peak and a trough,  Therefore, we split the analysis of states in the superposition into four cases: (1) Bob has a peak and Alice has a trough, (2) Alice has a peak and Bob has a trough, (3) Bob has both a peak and trough, and (4) Alice has both a peak and trough \footnote{In what follows, we assume that for cases (3) and (4) the peak and trough lie in different segments.  We may want to include an additional case where the peak and trough lie in the same segment, but this would require the existence of a segment of length $\geq L^2/2$.  Such a configuration only can occur on the 2D lattice if the entanglement cut intersects the labyrinth \emph{twice}, which is a measure zero instance of all possible labyrinth states}.  In cases (3) and (4), all of the segments associated to Alice and Bob respectively consist of sequences of open parentheses like $(((\cdots$ or closed parentheses like $)))\cdots$.  For instance, if Alice has a peak at segment $i$ and a trough at segment $j$, then the part of the wavefunction satisfying this constraint can be written as
\begin{equation}
 \ket{\psi_{(4), i, j}} = \left(\sum_{\substack{k,k' \\ |k'-k| = L^2/2}}\ket{\updiag}_{A_1}\ket{\updiag}_{A_2}\cdots \ket{\upbreak}_{A_i}\ket{\downdiag}_{A_{i+1}}\cdots\ket{\downbreak}_{A_j}\ket{\updiag}_{A_{j+1}}\cdots \right) \ket{\updiag}_{B_1}\ket{\updiag}_{B_2}\cdots \ket{\updiag}_{B_{i-1}}\ket{\downdiag}_{B_{i}}\cdots \ket{\updiag}_{B_{j}}\cdots
\end{equation}
where the state written above is purposely unnormalized and we have been slightly imprecise in the notation above, as the mountain region indicates configurations of Bell pairs (in the sense previously described above), which we have illustrated in the Figure below:
\begin{equation}
    \includegraphics[scale=0.5]{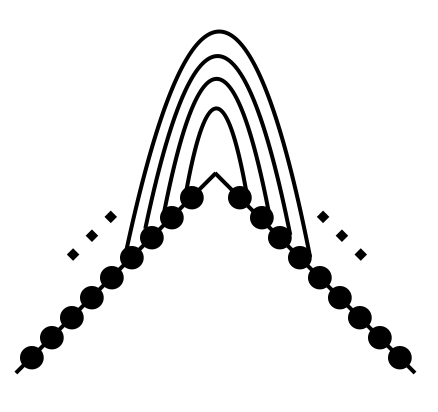}
\end{equation}
The sum in $\ket{\psi_{(4), i, j}}$ indicates a sum over the different locations for the mountain peak and mountain trough in segments $A_i$ and $A_j$ subject to the constraint that the peak and trough is separated by $L^2/2$ sites and remain in their respective segments.  A similar expression can be written down for $\ket{\psi_{(3), i, j}}$:
\begin{equation}
 \ket{\psi_{(3), i, j}} = \ket{\updiag}_{A_1}\ket{\updiag}_{A_2}\cdots \ket{\updiag}_{A_{i}}\ket{\downdiag}_{A_{i+1}}\cdots \ket{\updiag}_{A_{j+1}}\cdots\left(\sum_{\substack{k,k' \\ |k'-k| = L^2/2}}\ket{\updiag}_{B_1}\ket{\updiag}_{B_2}\cdots \ket{\upbreak}_{B_i}\ket{\downdiag}_{B_{i+1}}\cdots\ket{\downbreak}_{B_j}\ket{\updiag}_{B_{j+1}}\cdots \right). 
\end{equation}
Next, consider the state
\begin{equation}
    \ket{\psi_{4}} \propto \sum_{i,j} \ket{\psi_{(4), i, j}}.
\end{equation}
Taking a partial trace over Bob's half, 
% in particular $\psi_{(4), A} = \Tr_{B}{\ketbra{\psi_{(4)}}{\psi_{(4)}}}$, 
we find that 
\begin{equation}
    \psi_{(4), A} = \Tr_{B}{\ketbra{\psi_{(4)}}{\psi_{(4)}}} \propto \sum_{i,j} \Tr_B{\ketbra{\psi_{(4), i, j}}{\psi_{(4), i, j}}},
\end{equation}
which follows from the fact that Bob's subspace of states is disjoint for different values of $(i,j)$ for the peak and trough locations.  Similarly, 
\begin{equation}
    \psi_{(3), A} = \Tr_{B}{\ketbra{\psi_{(3)}}{\psi_{(3)}}} \propto \sum_{i,j} \Tr_B{\ketbra{\psi_{(3), i, j}}{\psi_{(3), i, j}}}.
\end{equation}
Next we turn to cases (1) and (2).  Here, the peak is in one of Alice's segments and the trough is in one of Bob's segments (or vice versa).  In this case, the wavefunction subject to these constraints can be written as
\begin{align}
 \ket{\psi_{(1), i, j}} &\propto \sum_{\substack{k,k' \\ |k'-k| = L^2/2}} \left(\ket{\updiag}_{A_1}\ket{\updiag}_{A_2}\cdots \ket{\upbreak}_{A_{i}}\ket{\downdiag}_{A_{i+1}}\cdots \ket{\updiag}_{A_{j+1}}\cdots\right)\left(\ket{\updiag}_{B_1}\ket{\updiag}_{B_2}\cdots \ket{\updiag}_{B_{i-1}}\ket{\downdiag}_{B_{i}}\cdots\ket{\downbreak}_{B_j}\ket{\updiag}_{B_{j+1}}\cdots \right) \nonumber\\
 &\equiv \sum_{\substack{k,k' \\ |k'-k| = L^2/2}} \ket{\phi^{(i,j)}_{(A,k),(B,k')}}.
\end{align}
and similarly
\begin{align}
 \ket{\psi_{(2), i, j}} &\propto \sum_{\substack{k,k' \\ |k'-k| = L^2/2}} \left(\ket{\updiag}_{A_1}\ket{\updiag}_{A_2}\cdots \ket{\updiag}_{A_{i}}\ket{\downdiag}_{A_{i+1}}\cdots\ket{\downbreak}_{A_j}\ket{\updiag}_{A_{j+1}}\cdots \right)\left(\ket{\updiag}_{B_1}\ket{\updiag}_{B_2}\cdots \ket{\upbreak}_{B_{i}}\ket{\downdiag}_{B_{i+1}}\cdots \ket{\updiag}_{B_{j+1}}\cdots\right) \nonumber \\
 &= \sum_{\substack{k,k' \\ |k'-k| = L^2/2}} \ket{\chi^{(i,j)}_{(A,k),(B,k')}} \equiv \sum_{\substack{k,k' \\ |k'-k| = L^2/2}} \ket{\phi^{(i,j)}_{(B,k),(A,k')}}.
\end{align}
As before, we can define $\ket{\psi_{(1)/(2)}} = \sum_{i,j} \ket{\psi_{(1)/(2),i,j}}$.  When computing the entanglement entropy, a helpful simplification occurs; because Bob's half of the state is disjoint for different $i,j,k,k'$, we find that
\begin{equation}
    \psi_{(1),A} = \frac{1}{\mathcal{N}_1}\sum_{i,j}\sum_{\substack{k,k' \\ |k'-k| = L^2/2}} \Tr_A\left(\ketbra{\phi^{(i,j)}_{(A,k),(B,k')}}{\phi^{(i,j)}_{(A,k),(B,k')}}\right)
\end{equation}
and
\begin{equation}
    \psi_{(2),A} = \frac{1}{\mathcal{N}_2}\sum_{i,j}\sum_{\substack{k,k' \\ |k'-k| = L^2/2}} \Tr_A\left(\ketbra{\chi^{(i,j)}_{(A,k),(B,k')}}{\chi^{(i,j)}_{(A,k),(B,k')}}\right)
\end{equation}
where $\mathcal{N}_{1,2}$ are unimportant normalization constants ensuring that the density matrix has unit trace.  It then follows (given that Bob's states are disjoint for all cases (1), (2), (3), (4)) the total entanglement entropy is
\begin{align}
    \psi_A =  p_{1} \psi_{(1),A} + p_2 \psi_{(2),A} + p_3\mathcal{N}_{3}^{-1}\sum_{i,j} \Tr_B{\ketbra{\psi_{(3), i, j}}{\psi_{(3), i, j}}} + p_4\mathcal{N}_{4}^{-1}\Tr_B{\ketbra{\psi_{(4), i, j}}{\psi_{(4), i, j}}}
\end{align}
where $p_{1,2}$ and $p_{3,4}$ are the probabilities of encountering configurations of cases (1) and (2) and cases (3) and (4) respectively (satisfying $p_1+p_2+p_3+p_4 = 1$), and $\mathcal{N}_{3,4}$ are unimportant normalization constants that ensure that density matrices have unit trace.  Since the reduced density matrices corresponding to the four cases have disjoint support, the entanglement entropy takes a rather simple form.  Indeed, we find 
\begin{equation}
    S_A(\psi) = p_{1}\mathbb{E}_{(i,j,k,k')}\left[S_A(\phi^{(i,j)}_{k,k'})\right] + p_{2}\mathbb{E}_{(i,j,k,k')}\left[S_A(\chi^{(i,j)}_{k,k'})\right] + p_{3} S_3 + p_4 S_4 - \sum_{m = 1,2,3,4} p_m \log p_m.
\end{equation}
where the term $p_{3} S_3 + p_4 S_4$ corresponds to the contribution from cases (3) and (4), which we ignore as these are harder to analyze and it is sufficient to consider the first two cases.  The expectation value is over $(i,j,k,k')$ subject to $|k-k'| = L^2/2$; this distribution is uniform over all quadruples satisfying the constraint.  The last term is a constant.

We first assume that $p_{1,2}$ are the same order of magnitude as $p_{3,4}$, which is a reasonable assumption for a randomly chosen partition of the loop into segments.  Next, we compute  $S_A(\phi^{(i,j)}_{k,k'})$ for a fixed value of $(i,j,k,k')$.  We note that the state $\phi$ in this case is a product state of Bell pairs organized into a mountain configuration; the entanglement entropy is simply the number of Bell pairs that have one half in the $A$-type segments and the other half in the $B$-type segments.  For a randomly chosen location for the mountain peak and partition, the number of such Bell pairs is roughly $O(L^2)$.  This value can become much smaller than $L^2$ only due to fine tuning the partition to be symmetric about the center of the mountain.  By symmetry, the same is true for the $\chi$ type states.  Therefore, the entanglement entropy is $O(L^2)$ for this toy problem.

In the above toy problem, we computed the entanglement entropy of a Motzkin chain decorating a single loop of length $L^2$, assuming the entanglement cut was chosen to intersect the loop a large number of times.  We note that it is only required to cut the loop at least twice to incur volume law entanglement so long as the total length of intervals assigned to Alice or Bob are roughly of similar length scales.  However, the true ground state is a superposition of all possible loop configurations decorated by Motzkin chains.  In particular, the ``labyrinth'' configuration can be drawn in many different ways on the lattice.  The full density matrix will thus contain off-diagonal terms that connect different loop configurations. Understanding how to account for these terms requires making use of slightly more heuristic methods.  

We first restrict the ground state wavefunction to loop configurations where the decorating Motzkin chain contains only single large-mountain configurations.  This restricted ground state wavefunction is given by
\begin{equation}
\ket{\psi} = \frac{1}{\sqrt{Z}}\sum_{C'} t^{A(C')} u^{\sum_{\ell \in L(C')} |\ell|^2} \ket{C'}.
\end{equation}
where $C'$ is a configuration of decorated loops subject to the above restriction, and the $\sum_{\ell \in L(C')} |\ell|^2$ appears because the mountain state has volume $|\ell|^2$ on a loop $\ell$ of length $|\ell|$.  Due to the Motzkin chain decorating each loop, we will adopt a pictorial way of describing this Motzkin chain, inspired by the toy calculation we did above.  

For each loop, we draw a directed colored line between each pair of matched parentheses.  This assignment of lines is unique, and each line corresponds to a pair of matched parentheses of a particular color (the directed property of the line indicates that the line points from the open to the closed parenthesis).  For a given configuration $C$, we can identify $L(I_{\partial A})$, viz. the set of loops which cross the boundary between $A$ and $B$.  For such a set of loops, define $\mathcal{B}(L)$ to be a map from the configurations decorating the loops to a set of ``matching lines'' connecting matched pairs of colored parentheses.  Call  $\mathcal{B}_{AB}(L) \subseteq \mathcal{B}(L)$ the subset of such pairs whose matching line crosses the boundary $\partial A$.  Then, define the normalized wavefunction
\begin{equation}
\ket{\psi_{\vec{b}}} = \frac{1}{\sqrt{Z_{\vec{b}}}} \sum_{C': \mathcal{B}_{AB}(L(C')) = \vec{b}} t^{A(C')} u^{\sum_{\ell \in L(C')} |\ell|^2}\ket{C'},
\end{equation}
which is the weighted superposition of all configurations satisfying the constraint that the ordered set of matching lines connecting from $A$ or $B$, notated as $\mathcal{B}_{AB}(L(C'))$, equals $\vec{b}$, which is an ordered vector indicating a colored sequence of matched lines \footnote{Here the ordering is induced by the orientation of a given loop.}.  We call $|\mathcal{B}_{AB}(L(C'))| = b$ and $\vec{b} \in \{0,1,2,\cdots, c-1\}^b$ where $c$ is the number of available colors.  Note that $Z_{\vec{b}} = Z_b$ is the same for all $\vec{b} \in \{0,1,2,\cdots, c-1\}^b$.  We may now write the full ground state as
\begin{equation}
    \ket{\psi} = \sum_b\sum_{\vec{b} \in \{0,1\}^b} \sqrt{\frac{Z_{\vec{b}}}{Z}} \ket{\psi_{\vec{b}}}
\end{equation}

We proceed by noting the following result.  Defining $\rho_A(\vec{b}) = \Tr_B(\ketbra{\psi_{\vec{b}}}{\psi_{\vec{b}}})$, we have that $\rho_A(\vec{b})$ and $\rho_A(\vec{b}')$ have different support if $\vec{b} \neq \vec{b}'$.  This in particular implies that $\Tr_B(\ketbra{\psi_{\vec{b}}}{\psi_{\vec{b}'}}) = 0$ for $\vec{b} \neq \vec{b}'$.  This follows from the fact that the Motzkin chains decorating each loop exactly form a mountain state. For more general configurations (with multiple peaks),  this is not true.  A simple explanation is to consider a configuration of parentheses on a fragment of a loop shown below:
\begin{equation}\label{eqn:subtlety}
    \includegraphics[scale=0.35]{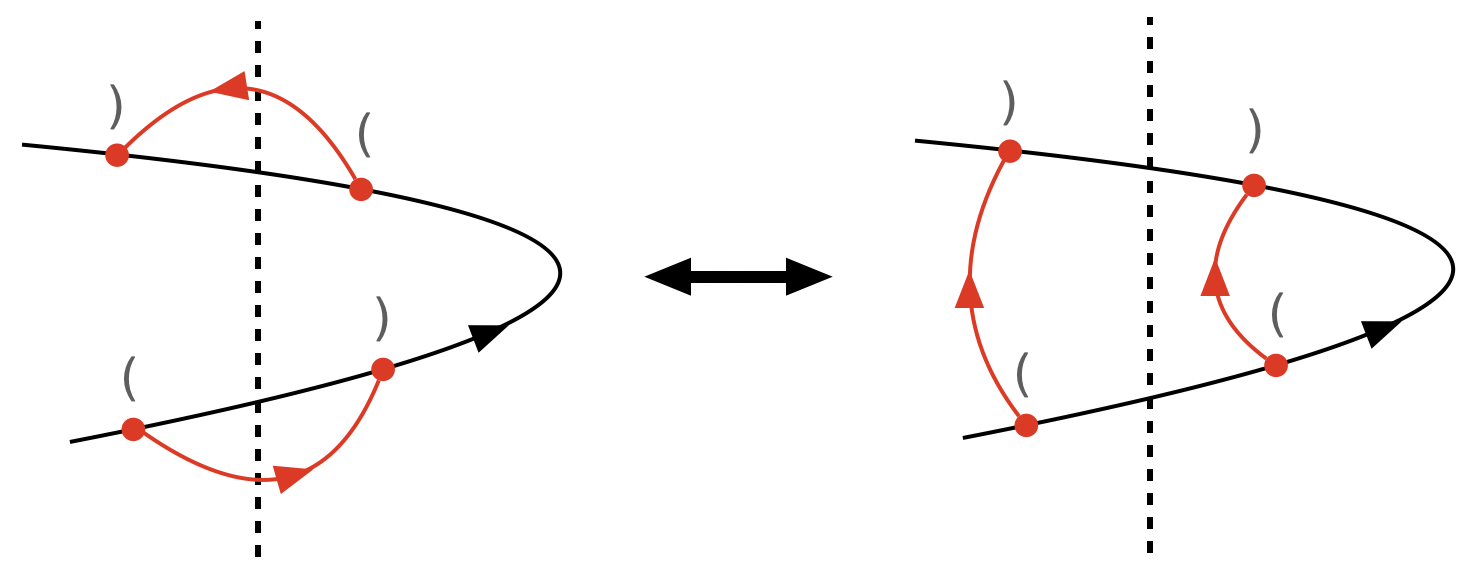}
\end{equation}
Rearranging Bell pairs as shown in the Figure above (which can be done while preserving the Motzkin constraint) does not change the configuration on the left half of the cut (indicated by the vertical dashed line) but the value of $\vec{b}$ has changed after the rearrangement to $\vec{b}'$.  This indicates that $\rho_A(\vec{b})$ and $\rho_A(\vec{b}')$ do not have orthogonal support.  However, the rearrangement of Bell pairs above changes the number of local extrema in the Motzkin configuration, which therefore means it cannot map mountain configurations to mountain configurations.  Since the above move cannot be made while staying in the space of single-mountain configurations, the disjoint support condition holds. Computing the entanglement entropy of this state, we find
\begin{equation}
    \rho_A = \Tr_B \ketbra{\psi}{\psi} = \sum_b \sum_{\vec{b} \in \{0,1\}^b} \frac{Z_{\vec{b}}}{Z} \Tr_B \ketbra{\psi_{\vec{b}}}{\psi_{\vec{b}}},
\end{equation}
and the entanglement entropy is 
\begin{equation}
    S(\rho_A) = \sum_{b} \sum_{\vec{b} \in \{0,1\}^b} \frac{Z_{\vec{b}}}{Z} S(\rho_A(\vec{b})) - \sum_b \sum_{\vec{b} \in \{0,1\}^b} \frac{Z_{\vec{b}}}{Z} \log \frac{Z_{\vec{b}}}{Z}.
\end{equation}
Notably, in the second term, we see that $Z_{\vec{b}}$ is the same for all $\vec{b}$ with the same length $b$, and therefore
\begin{equation}
    S(\rho_A) = \sum_{b} \sum_{\vec{b} \in \{0,1\}^b} \frac{Z_{\vec{b}}}{Z} S(\rho_A(\vec{b})) - \sum_b c^b \frac{Z_{b}}{Z} \log \frac{Z_{b}}{Z}.
\end{equation}
We define the probability distribution
\begin{equation}
    p(b) = c^b \frac{Z_b}{Z},
\end{equation}
so that
\begin{equation}
    S(\rho_A) = \sum_{b} \sum_{\vec{b} \in \{0,1\}^b} \frac{Z_{\vec{b}}}{Z} S(\rho_A(\vec{b})) - \sum_b p(b) \log p(b) + (\log c) \sum_b b \cdot p(b).
\end{equation}
Identifying the last term as proportional to $\langle b \rangle$, the entanglement entropy is thus lower bounded by the expected number of Bell pairs crossing from region $A$ to $B$ (the last term in the above equation).  Now, under the full distribution of loops, we know that the dominant contribution to this expectation value comes from the labyrinth states that contain a single large loop.  Indeed, with the decorated Motzkin chain, the total probability assigned to configurations with loops whose maximum loop length is $L^2 - \delta$ for constant $\delta$ is 
\begin{equation}
    \mathbb{P}(\max_{L; \, \ell \in L} |\ell| \leq L^2 - \delta) \leq (\text{dim}\, \mathcal{H})^{2 L^2} u^{(L^2 - \delta)^2/2 - L^2/2} \leq O(1)\cdot(u^{-\delta} \text{dim}^2\, \mathcal{H})^{L^2},
\end{equation}
which for large enough $\delta$ decays exponentially in $L^2$ \footnote{Here $(\dim \mch)^{2L^2}$ is a (rather loose) upper bound on the number of loop configurations that satisfy this property.}.  This is a rather drastic tail estimate since decreasing the length of the largest loop by a constant amount leads to a large suppression in amplitude.  This means that it is reasonable to focus our attention to loop configurations with a loop of length at least $L^2-\delta$, which from a previous argument has $O(L^2)$ Bell pairs on average crossing $\partial A$ for a random partition into two rectangular regions.  Therefore, this implies that $S(\rho_A) \geq O(L^2)$.

\medskip 

Note that we have not strictly speaking given a rigorous proof of the $O(L^2)$ estimate for the entanglement entropy, since in principle states not corresponding to mountain configurations should also be accounted for. We now sketch an argument to show that these configurations are unimportant.

Suppose we unravel an $L^2$ length loop to a straight line, and the mountain peak of the loop occurs at location $L^2/2$.  Then, if we were to deviate from the mountain configuration by reducing the height of the configuration at location $(1/2 - \epsilon) L^2$, for instance, by replacing a ``$($'' with a ``$0$'', the amplitude of the resulting configuration is suppressed by an amount $u^{-O(\epsilon L^2)}$ from the largest amplitude.  As a result, for $u$ large enough (set by $\text{dim}\,\mathcal{H}$), there exists an $\epsilon >0$ such that all typical configurations are caused by deviations of the height in some interval $\pm \epsilon L^2$ of the mountain peak.  Thus, the Bell pairs located in the region outside of the $\pm \epsilon L^2$ interval around the mountain peak remain untouched, and we may replace $\vec{b}$ by a vector of Bell pairs conditioned on being at a location outside of an interval $\pm \epsilon L^2$ around the mountain peak, and restricting the ground state to a superposition only over typical configurations.  In doing so, the number of Bell pairs in the region outside of the interval $\pm \epsilon L^2$ around the mountain peak is still $O(L^2)$ and thus we expect volume law entanglement through a very similar calculation.  However this argument relies on the fact that $u$ is larger than some constant set by $\text{dim}\,\mathcal{H}$, and does not guarantee a volume law for all $u > 1$ (which on physical grounds we expect to be the case).  There is likely an argument that the volume law phase persists for all $u > 1$, but we will not attempt to construct one here.

\subsection{Entanglement scaling when $t>1,u = 1$}\label{app:u1entang}

In this section we analyze the bipartite entanglement entropy of the decorated loop model when $u = 1$. In this case, different ground states $\bigotimes_{\ell} \ket{\Omega_{k_{\ell}}}$ can be formed corresponding to superpositions of all parenthetical strings with $k_{\ell}$ excess open or closed parentheses on loop $\ell$ (the ground state on $\ell$ is denoted by $\ket{\Omega_{k_{\ell}}}$).  Therefore, for sectors with any nonzero number of imbalanced parentheses on a loop, such loops will have restricted mobility, and cannot shrink into the vacuum state 
In the analysis below, we only consider the ergodic sector connected to the vacuum, i.e. the state where all loops are decorated by parenthetically balanced strings.  The ground state in this sector looks like
\begin{equation}
\ket{\psi} = \frac{1}{\sqrt{Z}} \sum_{C} t^{A(C)} \bigotimes_{\ell \in L(C)} \sqrt{Z_{\ell}(u=1)} \ket{\psi_{\ell}}
\end{equation}
where $\ket{\psi_{\ell}}$ is the unweighted Motzkin ground state on loop $\ell$.  

We now understand what the typical configurations look like.  Since when $u = 1$, loops no longer have a large entropy coming from their ``internal'' Motzkin degrees of freedom, typical configurations will be dominated by that of the biased colored loop model, rather than by the single-loop ``labyrinth'' configurations of the previous section.  Therefore, the dominant loop configurations are those that form the nested mountain shapes referred to in the main text, with such configurations possessing an amplitude of $O(t^{O(L^3)})$ \footnote{Note that here ``mountain'' refers to a mountain of loops, rather than a mountain formed by parenthesis on a particular loop (as in the previous section).}.  There are additional contributions to the amplitude coming from the internal Motzkin degrees of freedom, viz. $\prod_{\ell}\sqrt{Z_{\ell}(u=1)}$, but this is upper bounded by  $(\text{dim }\mathcal{H})^{O(L^2)}$ for constant $k$. For sufficiently large $t$, this additional contribution to the amplitude is therefore small in comparison to the $t>1$ weighting coming from the enclosed loop area.  

We now set out to compute the bipartite entanglement scaling.  We will provide a heuristic argument that the entanglement entropy is lower bounded by $O(L^{3/2})$.  We will first argue that there is a convenient decomposition for representing the ground state, which will aid us in computing the entanglement scaling.  

Let us assume that the ground state is a superposition of loop configurations where loops intersect the entanglement cut \emph{at most twice}.  This is true in the most probable configurations --- concentric nested loop mountains --- and also true upon deforming the loops mildly (apart from loops near the center of the mountain which are already small, but we discuss this case after the following analysis).  Pick a particular loop which intersects the entanglement cut twice.  Following the discussion in the previous appendix, we may assign ``matching'' lines which connect between parentheses that are matched for a given configuration of parentheses along the loop.  These matching lines are directed (pointing from open to closed parenthesis) and colored, and we only consider the set of such parentheses whose matching lines intersect with the entanglement cut.  A particular configurations of parentheses around a loop can be diagrammatically represented as
\begin{equation}
    \includegraphics[scale=0.35]{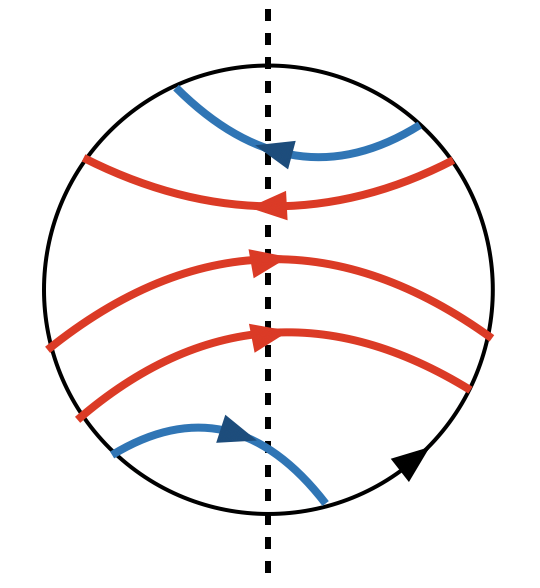}
\end{equation}
where we have only drawn matching pairs that cross the entanglement cut (dashed line), and the loop on which the parenthesis live corresponds to the outer black circle.  Note that a valid parenthetically matched configuration along the loop will always comprise of a sequence of matching lines pointing in one direction followed by a sequence of matching lines pointing in the opposite direction.  We now recall Eqn.~\ref{eqn:subtlety}, which provides a move between a pair of consecutive matching lines pointing in opposite directions such that the configuration in $A$ remains the same, while the configuration in $B$ changes.  This was used to argue that sectors corresponding to different numbers of matching lines are coupled together in the reduced density matrix.  Therefore, a decomposition of the reduced density matrix as
\begin{equation}
    \rho_A = \bigoplus_{\vec{\gamma}} \rho_{A, \vec{\gamma}}
\end{equation}
exists if $\vec{\gamma}$ corresponds to a vector of \emph{excess or deficit} matching lines.  In the example drawn above, we pair up neighboring matching lines until no matching lines can be paired any longer; this corresponds to:
\begin{equation}
    \includegraphics[scale=0.35]{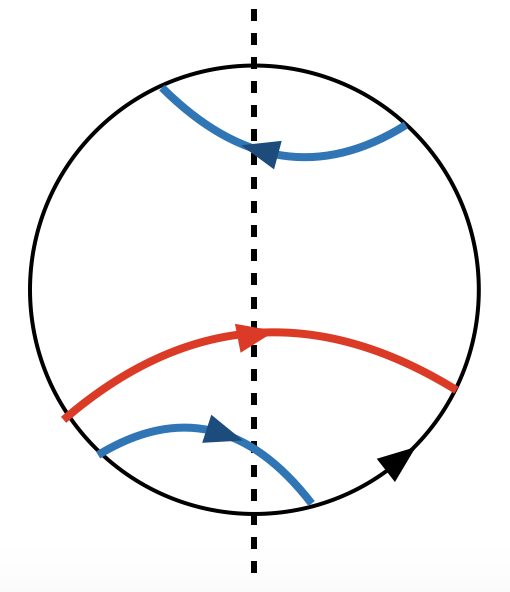}
\end{equation}
which in schematic notation gives $\gamma = \{\textcolor{blue}{\leftarrow}, \textcolor{red}{\rightarrow}, \textcolor{blue}{\rightarrow}\}$ \footnote{Incidentally this process of pairing up matching lines and eliminating them resembles a kind of parenthesis matching constraint.  However, we avoid using this terminology so as to avoid confusion with the other Motzkin constraints present in our models.}.  This ordered set of excess/deficit matching lines serves as a means to factorize the density matrix into blocks.  As a result, we may write the ground state wavefunction in the form: 
\begin{equation}
    \ket{\psi} = \sum_{\substack{I_{\partial A} \\ \vec{\gamma}_{\ell}: \ell \in L(I_{\partial A})}} \sqrt{\frac{Z_{AB}(I_{\partial A}, \vec{\gamma}_{\ell})}{Z}}\ket{\psi_{AB}(I_{\partial A}, \vec{\gamma}_{\ell})}
\end{equation}
where $\ket{\psi_{AB}(I_{\partial A}, \vec{\gamma}_{\ell})}$ is a uniform superposition over all configurations satisfying the desired boundary conditions along the cut.  Because of the factorizing structure of the density matrix, we may write
\begin{align}
    \psi_A &= \Tr_B \ketbra{\psi}{\psi} = \sum_{\substack{I_{\partial A} \\ \vec{\gamma}_{\ell}: \ell \in L(I_{\partial A})}} \frac{Z_{AB}(I_{\partial A}, \vec{\gamma}_{\ell})}{Z} \Tr_B \ketbra{\psi_{AB}(I_{\partial A}, \vec{\gamma}_{\ell})}{\psi_{AB}(I_{\partial A}, \vec{\gamma}_{\ell})} \nonumber \\
    &\triangleq \sum_{\substack{I_{\partial A} \\ \vec{\gamma}_{\ell}: \ell \in L(I_{\partial A})}} \frac{Z_{AB}(I_{\partial A}, \vec{\gamma}_{\ell})}{Z} \rho_{I_{\partial A}, \vec{\gamma}_{\ell}}.
\end{align}
Therefore, the entanglement entropy can be written as
\begin{equation}
S(\psi_A) = \sum_{\substack{I_{\partial A} \\ \vec{\gamma}_{\ell}: \ell \in L(I_{\partial A})}} \frac{Z_{AB}(I_{\partial A}, \vec{\gamma}_{\ell})}{Z} S(\rho_{I_{\partial A}, \vec{\gamma}_{\ell}}) - \sum_{\substack{I_{\partial A} \\ \vec{\gamma}_{\ell}: \ell \in L(I_{\partial A})}} \frac{Z_{AB}(I_{\partial A}, \vec{\gamma}_{\ell})}{Z} \log \frac{Z_{AB}(I_{\partial A}, \vec{\gamma}_{\ell})}{Z}.
\end{equation}
We note that changing colors in $\vec{\gamma}$ to $\vec{\gamma}'$ while preserving the number of excess/deficit directed matching lines does not affect the value of $Z_{AB}(I_{\partial A}, \vec{\gamma}_{\ell})$ or $S(\rho_{I_{\partial A}, \vec{\gamma}_{\ell}})$, and therefore, we can work with ``uncolored'' matching lines (denoted by $\vec{\overline{\gamma}}$) at the cost of multiplying by $c$ raised to an appropriate power; this gives 
\begin{equation}
S(\psi_A) = \sum_{I_{\partial A},\vec{\overline{\gamma}}_{\ell}} c^{\sum_{\ell} \overline{\gamma}_{\ell}} \frac{Z_{AB}(I_{\partial A}, \vec{\overline{\gamma}}_{\ell})}{Z} S(\rho_{I_{\partial A}, \vec{\overline{\gamma}}_{\ell}}) - \sum_{I_{\partial A},\vec{\overline{\gamma}}_{\ell}} c^{\sum_{\ell} \overline{\gamma}_{\ell}} \frac{Z_{AB}(I_{\partial A}, \vec{\overline{\gamma}}_{\ell})}{Z} \log \frac{Z_{AB}(I_{\partial A}, \vec{\overline{\gamma}}_{\ell})}{Z}.
\end{equation}
where 
\begin{equation}
    \overline{\gamma}_{\ell} \triangleq \begin{cases}
\norm{\vec{\overline{\gamma}}_{\ell}}_1 & \text{all } \leftarrow \text{ or all } \rightarrow \\
\norm{\vec{\overline{\gamma}}_{\ell}}_1-1+\frac{\log (c-1)}{\log c} & \text{otherwise}
\end{cases} 
\end{equation}
and the second case accounts for the fact that transition from a $\rightarrow$ to a $\leftarrow$ must come with a change in color, so the number of coloring options of the adjacent $\rightarrow \leftarrow$ pair is reduced from $c^2$ to $c^2-c$\footnote{$c^{\log(c-1)/\log(c) - 1} = 2^{\log(c-1) - \log(c)} = 1-1/c$, which produces the desired factor of $c(c-1)$ when paired with the contribution of the $\rightarrow\leftarrow$ pair to $\norm{\vec{\overline{\gamma}}_{\ell}}_1$.}.  In a similar vein to the previous calculations, we can define
\begin{equation}
    p(I_{\partial A}, \vec{\overline{\gamma}}_{\ell}) \triangleq c^{\sum_{\ell} \overline{\gamma}_{\ell}} \frac{Z_{AB}(I_{\partial A}, \vec{\overline{\gamma}}_{\ell})}{Z}
\end{equation}
so that
\begin{equation}
S(\psi_A) = \sum_{I_{\partial A},\vec{\overline{\gamma}}_{\ell}} p(I_{\partial A}, \vec{\overline{\gamma}}_{\ell}) S(\rho_{I_{\partial A}, \vec{\overline{\gamma}}_{\ell}}) - \sum_{I_{\partial A},\vec{\overline{\gamma}}_{\ell}} p(I_{\partial A}, \vec{\overline{\gamma}}_{\ell}) \log p(I_{\partial A}, \vec{\overline{\gamma}}_{\ell}) + \log c \left\langle \sum_{\ell} \overline{\gamma}_{\ell} \right\rangle.
\end{equation}
The second term gives factor of $O(L \log L)$ at most.  This follows from the fact that the total number of intersection points $I_{\partial A}$ is upper bounded by $(\text{dim }\mathcal{H})^{|\partial A|}$ and the total number of uncolored matching line assignments is upper bounded by $(L^2)^{|\partial A|}$;  therefore, the second term is upper bounded by $\log\left((\text{dim }\mathcal{H})^{|\partial A|}(L^2)^{|\partial A|}\right) = O(L \log L)$  The third term is the expected value of the number of deficit/excess matching lines.  We have to characterize typical configurations in order to proceed to give an approximate value for this quantity. 

Consider the value of $\varphi(L/2, L/2)$, viz. the number of loops encircling the center of the lattice.  Suppose that $\varphi(L/2, L/2) = L/2 - \delta$ for some $\delta$.  $\delta = 0$ would correspond to the single mountain configuration maximizing the enclosed volume.  If this is the case, then the maximum possible volume for such a configuration is $V_{\text{max}} - O(\delta^3)$.  As a result, the probability that $\varphi(L/2, L/2) \leq L/2 - \delta$ is
\begin{equation}
    \mathbb{P}(\varphi(L/2, L/2) \leq \alpha L/2) \leq (\text{dim }\mathcal{H})^{2L^2}\frac{t^{V_{\text{max}} - O(\delta^3)}}{t^{ V_{\text{max}}}} \leq t^{- O(L^2)}
\end{equation}
if $\delta = \alpha L^{2/3}$ for $\alpha$ large enough (see previous Appendix, where we used this tail bound to more formally argue that the TEE is anomalous in the biased colored loop model).  Since $\varphi(L/2, L/2)$ counts the number of loops surrounding the center of the lattice, there must be at least $L/2 - \delta$ loops crossing the entanglement cut, with $O(L)$ loops having length $O(L)$.  We expect the deficit/excess number of matching lines for each loop of length $O(L)$ to be $O(\sqrt{L})$.  Therefore, we expect $\left\langle \sum_{\ell} \overline{\gamma}_{\ell} \right\rangle$ to concentrate around $O(L^{3/2})$.  While we  at present cannot identify an expected scaling for the first term $\sum_{I_{\partial A},\vec{\overline{\gamma}}_{\ell}} p(I_{\partial A}, \vec{\overline{\gamma}}_{\ell}) S(\rho_{I_{\partial A}, \vec{\overline{\gamma}}_{\ell}})$, this argument at least places a lower bound on the entanglement entropy of $O(L^{3/2})$.   

Now, we address the assumption that the entanglement cut intersects each loop twice.  If we were to relax this assumption to intersecting some $O(1)$ number of times, we do not expect the scaling to change.  However, there is a possibility that the entanglement scaling can be dominated by labyrinth configurations, which have an extensive number of intersections (and are highly entangled).  To rule out this possibility, note that since the volume of typical configurations is $V_{\text{max}} - O(\delta^3)$ where $\delta = O(L^{2/3})$, there is at most a region of area $O(L^{4/3})$ on the lattice (around the center of the mountain) where loop configurations can, in principle, be unconstrained.  However, the most entropy we can pack in this region scales like the size of the region, or $O(L^{4/3})$, which cannot dominate over the $O(L^{3/2})$ scaling.  

\begin{figure}
    \centering
    \includegraphics[scale=0.18]{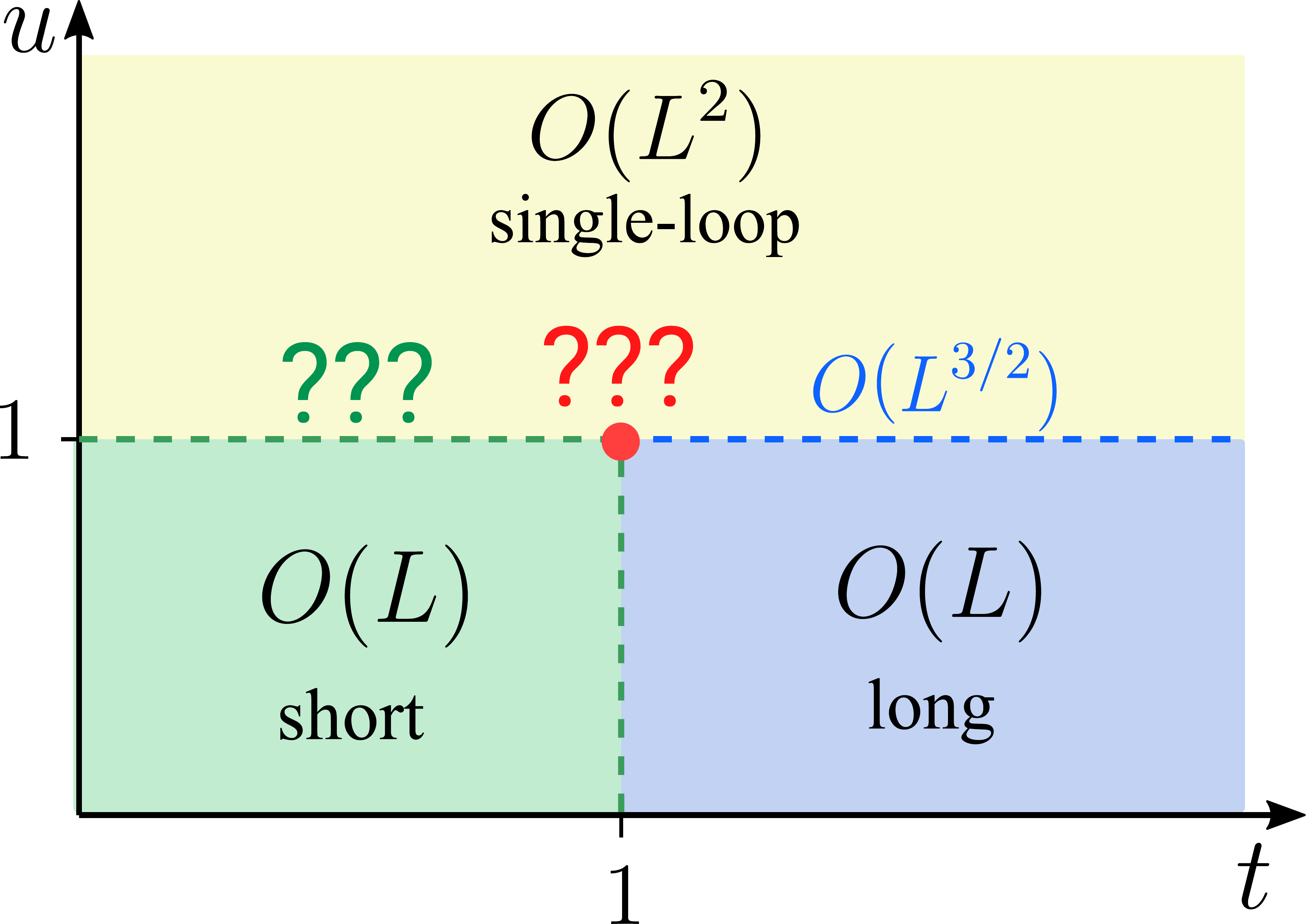}
    \caption{Purported phase diagram for the decorated loop model.  Dashed blue lines indicate entanglement transitions, and dotted green lines indicate quantum phase transitions which may correspond to entanglement transitions.  The multicritical point is shown in red.}
    \label{fig:groundstate}
\end{figure}

\subsection{General phase diagram of the decorated model}\label{app:phase_diagram}

We now present a general phase diagram illustrating the various phases and entanglement scaling in the decorated model:
\begin{enumerate}
\item $u > 1$: Regardless of the value of $t$, the bipartite entanglement entropy will exhibit volume law scaling.  This is because the dominant contribution to the ground state features a single loop (or $O(1)$ loops) each of length $O(L^2)$.
\item $u < 1$, $t > 1$: In this regime, loops are long, but the Motzkin chains decorating each loop have area law entanglement \cite{klich}.  Therefore, we expect area law entanglement scaling.  If the number of colors of loops (rather than colors of decorating parentheses) satisfied $|c| > 1$, this phase would have anomalous TEE.
\item $u < 1$, $t < 1$: In this regime, loops are short, so we expect area law entanglement.  In general, we expect this phase to be adiabatically connected to a trivial gapped phase.
\item $u = 1$, $t > 1$: Loops remain long in this regime, but now an unweighted Motzkin chain decorates the loops.  Since the unweighted Motzkin chain has $\sqrt{L}$ entanglement entropy, this critical line has $O(L^{3/2})$ entanglement scaling.
\end{enumerate}
It seems likely that the line $u < 1$, $t = 1$ indicates a transition between the long loop and short loop phases; this follows from the fact that the height field fluctuations follow a free field without a source term (which otherwise provides a volume-deforming modification to the wavefunction amplitudes):
\begin{equation}\label{eq:cqcp}
\ket{\psi} \approx \frac{1}{\sqrt{Z}} \int \mathcal{D} \varphi \, \Theta(\varphi)\, \exp(-\frac{1}{2}\int d^2x\, (\nabla \varphi)^2) \ket{\{\varphi\}}.
\end{equation}
where the constraint $\Theta(\varphi)$ ensures that the height field is always positive.  This notation follows Ref.~\cite{fradkinmoore}, which considered conformal quantum critical points; i.e. wavefunctions whose amplitudes are Boltzmann weights of a classical statistical mechanics model tuned to a critical point.  Examples of this include quantum dimer models on bipartite lattices and certain quantum loop models.

It was shown \cite{fradkinmoore} that wavefunctions corresponding to conformal quantum critical points have the entanglement scaling
\begin{equation}
    S(A) = \alpha |\partial A| - \frac{c}{6}(\chi_A + \chi_B - \chi_{A \cup B}) \log |\partial A|,
\end{equation}
where $\chi$ denotes the Euler characteristic.  Since the entanglement scaling possesses a subleading logarithmic correction, the TEE, which scales like $\log |\partial A|$, probes the transition between the two area law phases.  If there are more than two colors of loops, then the TEE is expected to exhibit anomalous TEE along the line $u < 1$, $t = 1$.

The nature of the entanglement scaling along $u = 1$, $t < 1$ is unknown: while loops prefer to be short, the entanglement along each loop is large (scaling like $O(\sqrt{\ell})$ with $\ell$ the length of the loop), and the entropic force due to this large entanglement may result in loops becoming large.  If this favors small loops, then we expect area law scaling, and if long loops are favored, we expect power law violations to this entanglement scaling.  More interesting is a scenario where the competition favors intermediate length loops, which could be a route to new and exotic entanglement scaling.   

For a similar reason, the entanglement scaling at the multicritical point $u = t = 1$ is unknown to us.  It is likely even larger than along $u = 1$, $t < 1$, because there is no longer a constraint forcing loops to be short.  It would be an interesting future direction to rigorously characterize the entanglement scaling, either by analytical or numerical means.

\subsection{The case when $|d| = 1$}\label{app:single_color_parenthesis}

Consider now the case with only a single color of decorating parenthesis. The analysis in this case is significantly different.  In particular, along the line $u = 1$, $t > 1$, an identical analysis to the one above gives an entanglement scaling of $O(L \log L)$.  This is because the entanglement entropy of a Motzkin chain with a {\it single} colored set of parentheses scales like $\log L$.  The analysis in this case is nearly identical to the one carried out for multiple colors, so we omit a proof. 

Moreover, for a single color of parenthesis we expect that the previously volume law phase, which was dominated by a single loop winding around the lattice and decorated by a colored Motzkin chain, now has different entanglement scaling since the decorating Motzkin chain no longer possesses large entanglement.  Given that the uncolored deformed Motzkin chain only has area law bipartite entanglement, we conjecture that this is an area law phase.  A full proof of this would be an important future direction.

\section{Decorated fully packed loop model}\label{app:decorated_packed_loop}

In the main text we discussed loop models where loops explicitly do not merge or split with each other.  It is useful to understand if one may formulate a similar model where the dominant exchange mechanism is loop splitting/merging.  This is for instance the case in fully-packed loop models.  In this section we thus consider a fully-packed version of the decorated model discussed above. 

Consider a trivalent lattice, which for simplicity we will choose to be the square-octagon lattice.  As before, we define a local Hilbert space of the form
\begin{equation}
\mathcal{H} = \mathcal{H}_{\text{vac}} \oplus (\mathcal{H}_{\text{arr}} \otimes \mathcal{H}_{\text{mot}})
\end{equation}
where $\mathcal{H}_{\text{vac}} = \text{span}\{\ket{-}\}$, $\mathcal{H}_{\text{arr}} = \text{span}\{\ket{\leftarrow}, \ket{\rightarrow}\}$ (there are four different link orientations, so this local Hilbert space is modified accordingly for each of the orientations), and 
\begin{equation}
\mathcal{H}_{\text{mot}} = \text{span}\{\ket{(_d}, \ket{)_d}, \ket{0}, \ket{\alpha}\}.
\end{equation}
The new state $|\alpha\rangle$ signifies the start/end of a sequence of matched parenthesis; its presence is needed because in order for two loops to merge and/or split, it is necessary to know when a sequence of parentheses has started or ended, otherwise any loop merging/splitting process agnostic to this can violate the Motzkin constraint. 

The Hilbert space constraint is imposed by $H_{\text{con}}$.  Note that the centers of the squares on the square-octagon lattice form a square superlattice, which we divide into $A$ and $B$ sublattices.  On the $A$ sublattice, we restrict to the dimer configurations
\begin{equation}
\begin{array}{cc}
  \begin{tikzpicture}[baseline={([yshift=1.5ex]current bounding box.center)},vertex/.style={anchor=base,circle,fill=black!25,minimum size=18pt,inner sep=2pt}]
    \draw[->, line width=0.2mm] (0.4,0)--(0,0);
    \draw[->, line width=0.2mm] (0.4*2.414,0)--(0.4*3.414,0);
    \draw[<-, line width=0.2mm] (0.4*1.707,0.4*0.707)--(0.4*1.707,0.4*1.707);
    \draw[<-, line width=0.2mm] (0.4*1.707,-0.4*0.707)--(0.4*1.707,-0.4*1.707);
    \draw[<-, line width=0.2mm] (0.4,0)--(0.4*1.707,0.4*0.707);
    \draw[<-, line width=0.2mm] (0.4*2.414,0)--(0.4*1.707,-0.4*0.707);
    \draw[-, dashed] (0.4,0)--(0.4*1.707,-0.4*0.707);
    \draw[-, dashed] (0.4*2.414,0)--(0.4*1.707,0.4*0.707);
  \end{tikzpicture} \hspace{0.5cm} & \begin{tikzpicture}[baseline={([yshift=1.5ex]current bounding box.center)},vertex/.style={anchor=base,circle,fill=black!25,minimum size=18pt,inner sep=2pt}]
    \draw[->, line width=0.2mm] (0.4,0)--(0,0);
    \draw[->, line width=0.2mm] (0.4*2.414,0)--(0.4*3.414,0);
    \draw[<-, line width=0.2mm] (0.4*1.707,0.4*0.707)--(0.4*1.707,0.4*1.707);
    \draw[<-, line width=0.2mm] (0.4*1.707,-0.4*0.707)--(0.4*1.707,-0.4*1.707);
    \draw[-, dashed] (0.4,0)--(0.4*1.707,0.4*0.707);
    \draw[-, dashed] (0.4*2.414,0)--(0.4*1.707,-0.4*0.707);
    \draw[<-, line width=0.2mm] (0.4,0)--(0.4*1.707,-0.4*0.707);
    \draw[<-, line width=0.2mm] (0.4*2.414,0)--(0.4*1.707,0.4*0.707);
  \end{tikzpicture}\\
\end{array}
\end{equation}
while on the $B$ sublattice, we restrict to the dimer configurations
\begin{equation}
\begin{array}{cc}
  \begin{tikzpicture}[baseline={([yshift=1.5ex]current bounding box.center)},vertex/.style={anchor=base,circle,fill=black!25,minimum size=18pt,inner sep=2pt}]
    \draw[<-, line width=0.2mm] (0.4,0)--(0,0);
    \draw[<-, line width=0.2mm] (0.4*2.414,0)--(0.4*3.414,0);
    \draw[->, line width=0.2mm] (0.4*1.707,0.4*0.707)--(0.4*1.707,0.4*1.707);
    \draw[->, line width=0.2mm] (0.4*1.707,-0.4*0.707)--(0.4*1.707,-0.4*1.707);
    \draw[->, line width=0.2mm] (0.4,0)--(0.4*1.707,0.4*0.707);
    \draw[->, line width=0.2mm] (0.4*2.414,0)--(0.4*1.707,-0.4*0.707);
    \draw[-, dashed] (0.4,0)--(0.4*1.707,-0.4*0.707);
    \draw[-, dashed] (0.4*2.414,0)--(0.4*1.707,0.4*0.707);
  \end{tikzpicture} \hspace{0.5cm} & \begin{tikzpicture}[baseline={([yshift=1.5ex]current bounding box.center)},vertex/.style={anchor=base,circle,fill=black!25,minimum size=18pt,inner sep=2pt}]
    \draw[<-, line width=0.2mm] (0.4,0)--(0,0);
    \draw[<-, line width=0.2mm] (0.4*2.414,0)--(0.4*3.414,0);
    \draw[->, line width=0.2mm] (0.4*1.707,0.4*0.707)--(0.4*1.707,0.4*1.707);
    \draw[->, line width=0.2mm] (0.4*1.707,-0.4*0.707)--(0.4*1.707,-0.4*1.707);
    \draw[-, dashed] (0.4,0)--(0.4*1.707,0.4*0.707);
    \draw[-, dashed] (0.4*2.414,0)--(0.4*1.707,-0.4*0.707);
    \draw[->, line width=0.2mm] (0.4,0)--(0.4*1.707,-0.4*0.707);
    \draw[->, line width=0.2mm] (0.4*2.414,0)--(0.4*1.707,0.4*0.707);
  \end{tikzpicture}\\
\end{array}
\end{equation}
Calling the set of these configurations $\mathscr{C}_{A,B}$, the constraint Hamiltonian can be defined as $H_{\text{con},A,s} = -\sum_{C \in \mathscr{C}_{A,s}} \ketbra{C}{C}$ and $H_{\text{con},B,s} = -\sum_{C \in \mathscr{C}_{B,s}} \ketbra{C}{C}$, where $s$ indicates a square on the $A$ or $B$ sublattice.  The total constraint Hamiltonian is $H_{\text{con}} = \sum_{s \in A} H_{\text{con},A,s} + \sum_{s \in B} H_{\text{con},B,s}$

Next we describe the exchange terms that toggle between local configurations.  We consider any pair of consecutive arrows, which for simplicity we will not draw out spatially in the lattice:
\begin{equation}
\begin{array}{c}
  \begin{tikzpicture}[baseline={([yshift=1.5ex]current bounding box.center)},vertex/.style={anchor=base,circle,fill=black!25,minimum size=18pt,inner sep=2pt}]
    \draw[-, line width=0.3mm] (0,0)--(1,0);
    \draw[->, line width=0.3mm] (0,0)--(0.5,0) node[above] {$c_1$};
     \draw[-, line width=0.3mm] (1.0,0)--(2.0,0);
    \draw[->, line width=0.3mm] (1.0,0)--(1.5,0) node[above] {$c_2$};
  \end{tikzpicture}
\end{array}
\end{equation}
We define the following processes on the decorated characters $c_1$ and $c_2$
\begin{equation}
H_{\text{kin}, v, 1}(u) = \mathcal{P}_v\left(\ket{(_d)_d} - u \ket{00}\right) + \mathcal{P}_v\left(\ket{(_d 0} - u \ket{0 (_d}\right) + \mathcal{P}_v\left(u \ket{0 )_d} - \ket{)_d 0}\right)
\end{equation}
We also have additional processes involving the new $\alpha$ character as we enforce the condition that the string of characters between consecutive $\alpha$'s forms a parenthetically matched sequence.  This can be enforced by
\begin{equation}
H_{\text{kin}, v, 2} = \mathcal{P}_v\left(\ket{(_d \alpha}\right) + \mathcal{P}_v\left(\ket{\alpha )_d}\right) + \mathcal{P}_v\left( \ket{0 \alpha} - \ket{\alpha 0}\right),
\end{equation}
where the last condition hops an $\alpha$-type character past a 0.  We have analogous exchange terms when both of the arrows are flipped (one must interchange the positions of $c_1$ and $c_2$).  Finally, we need a condition that merges loops while preserving the Motzkin constraint on each loop; we denote this by $H_{\text{kin}, v, 3}$.  The rule(s) which accomplish this is
\begin{equation}
  \begin{tikzpicture}[baseline={([yshift=-0.5ex]current bounding box.center)},vertex/.style={anchor=base,circle,fill=black!25,minimum size=18pt,inner sep=2pt}]
    \draw[->, line width=0.2mm] (0.4,0)--(0,0);
    \draw[->, line width=0.2mm] (0.4*2.414,0)--(0.4*3.414,0);
    \draw[<-, line width=0.2mm] (0.4*1.707,0.4*0.707)--(0.4*1.707,0.4*1.707);
    \draw[<-, line width=0.2mm] (0.4*1.707,-0.4*0.707)--(0.4*1.707,-0.4*1.707);
    \draw[<-, line width=0.2mm] (0.4,0)--(0.4*1.707,0.4*0.707) node[left] {$\alpha$};
    \draw[<-, line width=0.2mm] (0.4*2.414,0)--(0.4*1.707,-0.4*0.707) node[right] {$\alpha$};
    \draw[-, dashed] (0.4,0)--(0.4*1.707,-0.4*0.707);
    \draw[-, dashed] (0.4*2.414,0)--(0.4*1.707,0.4*0.707);
  \end{tikzpicture} \hspace{0.2cm} \Longleftrightarrow \hspace{0.2cm} \begin{tikzpicture}[baseline={([yshift=-0.5ex]current bounding box.center)},vertex/.style={anchor=base,circle,fill=black!25,minimum size=18pt,inner sep=2pt}]
    \draw[->, line width=0.2mm] (0.4,0)--(0,0);
    \draw[->, line width=0.2mm] (0.4*2.414,0)--(0.4*3.414,0);
    \draw[<-, line width=0.2mm] (0.4*1.707,0.4*0.707)--(0.4*1.707,0.4*1.707);
    \draw[<-, line width=0.2mm] (0.4*1.707,-0.4*0.707)--(0.4*1.707,-0.4*1.707);
    \draw[-, dashed] (0.4,0)--(0.4*1.707,0.4*0.707);
    \draw[-, dashed] (0.4*2.414,0)--(0.4*1.707,-0.4*0.707);
    \draw[<-, line width=0.2mm] (0.4,0)--(0.4*1.707,-0.4*0.707) node[left] {$\alpha$};
    \draw[<-, line width=0.2mm] (0.4*2.414,0)--(0.4*1.707,0.4*0.707) node[right] {$\alpha$};
  \end{tikzpicture}
\end{equation}
on the $A$ sublattice, while
\begin{equation}
  \begin{tikzpicture}[baseline={([yshift=-0.5ex]current bounding box.center)},vertex/.style={anchor=base,circle,fill=black!25,minimum size=18pt,inner sep=2pt}]
    \draw[<-, line width=0.2mm] (0.4,0)--(0,0);
    \draw[<-, line width=0.2mm] (0.4*2.414,0)--(0.4*3.414,0);
    \draw[->, line width=0.2mm] (0.4*1.707,0.4*0.707)--(0.4*1.707,0.4*1.707);
    \draw[->, line width=0.2mm] (0.4*1.707,-0.4*0.707)--(0.4*1.707,-0.4*1.707);
    \draw[->, line width=0.2mm] (0.4,0)--(0.4*1.707,0.4*0.707) node[left] {$\alpha$};
    \draw[->, line width=0.2mm] (0.4*2.414,0)--(0.4*1.707,-0.4*0.707) node[right] {$\alpha$};
    \draw[-, dashed] (0.4,0)--(0.4*1.707,-0.4*0.707);
    \draw[-, dashed] (0.4*2.414,0)--(0.4*1.707,0.4*0.707);
  \end{tikzpicture} \hspace{0.2cm} \Longleftrightarrow \hspace{0.2cm} \begin{tikzpicture}[baseline={([yshift=-0.5ex]current bounding box.center)},vertex/.style={anchor=base,circle,fill=black!25,minimum size=18pt,inner sep=2pt}]
    \draw[<-, line width=0.2mm] (0.4,0)--(0,0);
    \draw[<-, line width=0.2mm] (0.4*2.414,0)--(0.4*3.414,0);
    \draw[->, line width=0.2mm] (0.4*1.707,0.4*0.707)--(0.4*1.707,0.4*1.707);
    \draw[->, line width=0.2mm] (0.4*1.707,-0.4*0.707)--(0.4*1.707,-0.4*1.707);
    \draw[-, dashed] (0.4,0)--(0.4*1.707,0.4*0.707);
    \draw[-, dashed] (0.4*2.414,0)--(0.4*1.707,-0.4*0.707);
    \draw[->, line width=0.2mm] (0.4,0)--(0.4*1.707,-0.4*0.707) node[left] {$\alpha$};
    \draw[->, line width=0.2mm] (0.4*2.414,0)--(0.4*1.707,0.4*0.707) node[right] {$\alpha$};
  \end{tikzpicture}
\end{equation}
on the $B$ sublattice.  One may define $H_{\text{kin}, s, 3, A/B} = \mathcal{P}_{s}(\ket{M(0)} - \ket{M(1)})$ using these rules above, and $H_{\text{kin}, s, 3} = H_{\text{kin}, s, 3, A} + H_{\text{kin}, s, 3, B}$.  Defining $H_{\text{kin}, 1}(u) = \sum_v H_{\text{kin}, v, 1}(u)$, $H_{\text{kin}, 2} = \sum_v H_{\text{kin}, v, 2}$, and $H_{\text{kin}, 3} = \sum_s H_{\text{kin}, s, 3}$, the Hamiltonian
\begin{equation}
H(u) = H_{\text{con}} + H_{\text{kin}, 1}(u) + H_{\text{kin}, 2} + H_{\text{kin}, 3}
\end{equation}
is frustration free and has a ground state which is a uniform superposition of fully packed loops on the square-octagon lattice, where contiguous strings between consecutive $\alpha$'s form a Motzkin chain.  The kinetic term for the $\alpha$ characters allows them to hop across the chain, subject to the constraint described above.  However, there is a $U(1)$ constraint on the number of $\alpha$ type characters, which results in a large ground state degeneracy. The ground state sectors can be labelled by a subset $\mathcal{S}$ of ``frozen'' loops (which include no $\alpha$ characters); in the regions outside of those occupied by the loops in $\mathcal{S}$,  the ground states resemble those of a fully-packed loop model with loops decorated by a Motzkin chain between subsequent occurrences of $\alpha$.  The entanglement scaling of this model is a very interesting question that we leave to a future study.
\end{document}